\documentclass[12pt]{iopart}
\usepackage{graphicx}
\usepackage{dcolumn}
\makeatletter
\expandafter\let\csname equation*\endcsname\relax
\expandafter\let\csname endequation*\endcsname\relax
\makeatother
\usepackage{amsmath}
\usepackage{amssymb}
\usepackage{bm}
\usepackage{orcidlink}
\hypersetup{hidelinks}
\usepackage{subcaption}
\usepackage{tabularx}
\usepackage{float}
\begin{document}

\title{Directional Anisotropic Sensitivity Curves for Pulsar Timing Arrays}

\author{Daniel J. Oliver \orcidlink{0000-0002-7374-6925}$^{1}$, Jeffrey S. Hazboun \orcidlink{0000-0003-2742-3321}$^1$, Martine M. Maggi \orcidlink{0009-0002-7859-0023}$^{1}$, Jeremy G. Baier \orcidlink{0000-0002-4972-1525}$^1$, Kyle E. Gourlie \orcidlink{0000-0002-8554-3201}$^{1,2}$}

\address{$^1$ Oregon State University, 1500 SW Jefferson Ave, Corvallis, OR 97331, USA}
\address{$^2$ University of Wisconsin--Milwaukee, 2200 E Kenwood Blvd, Milwaukee, WI 53211, USA}

\ead{oliverda@oregonstate.edu}
\vspace{10pt}

\begin{abstract}
After two decades of observations, pulsar timing array collaborations have reported strong evidence for a stochastic gravitational wave background, most likely sourced by an inspiraling population of supermassive black hole binaries. Because that population is finite, anisotropy in the background is inevitable, producing hotspots on the sky that track the loudest binaries. Building on the Fisher formalism for anisotropic backgrounds, we recast the directional Fisher information as effective sensitivity curves on the sky, distinguishing the radiometer and full-Fisher estimators, along with a sky-weighted curve that reduces the anisotropic sky to a single spectrum and recovers the standard isotropic curve in the isotropic limit. We implement five sky-decomposition bases: pixel, spherical harmonic, square-root spherical harmonic, radiometer, and principal-map. We demonstrate the framework on a simulated array modeled on the International Pulsar Timing Array: characterizing the angular response, analyzing the Fisher structure, quantifying the detectability of injected anisotropic hotspots, and forecasting future-array sensitivity. The framework is released as an extension to the sensitivity software package \texttt{hasasia}.
\end{abstract}


\maketitle

\section{\label{sec:Intro}Introduction\protect}

Pulsar timing was first proposed as a gravitational wave (GW) detector in the 1970s by two independent groups \cite{sazhin_opportunities_1978,detweiler_pulsar_1979}, both of whom showed that a passing gravitational wave would imprint a small, frequency-dependent shift in the arrival times of pulses from a precisely timed pulsar. A decade later, Foster \& Backer (1990) \cite{foster_constructing_1990} extended this idea, proposing a coordinated array of millisecond pulsars timed regularly and analyzed jointly, forming the basis for the modern pulsar timing array (PTA).

In June 2023, PTA collaborations around the globe collectively announced evidence for the detection of a stochastic gravitational wave background (GWB) with Hellings-Downs (HD) \cite{hellings_upper_1983} spatial correlations: NANOGrav \cite{agazie_nanograv_2023}, EPTA/InPTA \cite{epta_collaboration_and_inpta_collaboration_second_2023}, PPTA \cite{reardon_search_2023}, and the CPTA \cite{xu_searching_2023}. The recovered amplitude $A_{\mathrm{GWB}} \sim 2.4 \times 10^{-15}$ at a reference frequency of $1/\mathrm{yr}$ is consistent with theoretical predictions for a population of supermassive black hole binaries (SMBHBs) \cite{phinney_practical_2001,sesana_stochastic_2008,sesana_systematic_2013}.

A PTA extracts gravitational wave information by cross-correlating the timing residuals from many pulsars \cite{romano_detection_2017}. A stochastic GWB imprints a characteristic angular pattern (the HD correlation) on every pulsar pair that distinguishes it from any intrinsic pulsar noise. Standard PTA analyses assume the GWB is isotropic on the sky and average the array's directional sensitivity over the sky \cite{arzoumanian_nanograv_2020}. Both of these are working approximations, and both are relaxed with the directional framework developed in this paper.

The SMBHB population that sources this background is built up through hierarchical galaxy mergers \cite{begelman_massive_1980, rajagopal_ultra-low_1995, jaffe_gravitational_2003}. The central black holes in massive galaxies follow well-known scaling relations with their hosts' stellar properties \cite{kormendy_inward_1995,ferrarese_fundamental_2000, gebhardt_relationship_2000, mcconnell_revisiting_2013, kormendy_coevolution_2013}, and the demographics implied by these relations are broadly consistent with the observed GWB amplitude (see \cite{burke-spolaor_astrophysics_2019} for a review).

Because this population is finite, the resulting GWB cannot be perfectly isotropic. The brightest and nearest binaries should generate hotspots in the angular power distribution on the sky \cite{mingarelli_characterizing_2013, mingarelli_local_2017}, with the expected anisotropy growing with frequency as $C_\ell/C_0 \propto f^{11/3}$ in the gravitational-wave-driven regime \cite{sato-polito_exploring_2024, lin_analytical_2026, raidal_2026}. Detection prospects have been quantified by \cite{lemke_detecting_2024}. A hotspot may be a single loud SMBHB or the sum of several quieter binaries close together on the sky. PTA collaborations have already searched for anisotropy with no significant detection to date, placing upper limits on $C_\ell/C_0$ \cite{taylor_limits_2015, agazie_nanograv_2023-2, grunthal_optimising_2026, chen_searching_2026}. 

There are two types of anisotropy in PTAs, and the two must be disentangled to fully characterize the GWB. The first is astrophysical: the GWB itself is anisotropic at some level with hotspots in the directions of the loudest binaries. The second is instrumental: pulsars are not uniformly distributed on the sky, and general relativity's quadrupolar antenna pattern delivers very different directional sensitivity in pulsar-dense and pulsar-sparse regions. We refer to this second component as the array's \emph{angular response}. Without a quantitative model of the angular response, signal and detector anisotropy cannot be separated.

Sensitivity curve tools for PTAs are well developed in the isotropic regime \cite{hazboun_realistic_2019, hazboun_hasasia_2019, baier_sensitivity_2025}, building on the analytic sensitivity literature for ground- and space-based stochastic-background searches \cite{thrane_probing_2009, moore_gravitational-wave_2015, romano_detection_2017} and sky-averaged detection forecasts \cite{taylor_are_2016}; none of these resolve sensitivity by direction. For anisotropy specifically, a cross-correlation Fisher formalism was developed by Ali-Ha\"imoud, Smith, \& Mingarelli (2020, 2021) \cite{ali-haimoud_fisher_2020, ali-haimoud_insights_2021} and applied in the pixel basis by Pol, Taylor, \& Romano (2022) \cite{pol_forecasting_2022}, primarily to forecast detectability and constrain the angular power $C_\ell/C_0$. The \texttt{fastPTA} package \cite{babak_forecasting_2024, depta_pulsar_2025} likewise forecasts multipole-level sensitivity, a sky-integrated rather than sky-resolved quantity. Most closely related, Moursy et al. (2026) \cite{moursy_anisotropy_2026} construct anisotropic sensitivity curves empirically, calibrating scaling relations and a NANOGrav forecast from large simulation suites. What remains missing is an analytic directional effective sensitivity $S_{\mathrm{eff}}(f,\hat{\Omega})$ for arbitrary anisotropic sky power, derived directly from per-pulsar noise models. 

This paper fills that gap. We extend \texttt{hasasia}\footnote{https://github.com/Hazboun6/hasasia} with an analytic framework for directional effective sensitivity, derived in closed form from its per-pulsar noise models, and implement it in five sky-decomposition bases: pixel, spherical harmonic, square-root spherical harmonic, radiometer, and principal-map. We demonstrate the framework on a simulated International Pulsar Timing Array (IPTA)-like array, characterizing its angular response, quantifying the detectability of injected anisotropic hotspots, and forecasting its scaling to future array configurations. 

The remainder of this paper is organized as follows. \S\ref{sec:methods} develops the directional Fisher framework starting from the cross-correlation estimator, including the translation to the Fourier domain, the directional effective sensitivity $S_{\mathrm{eff}}$, and the connection to PTA map making. \S\ref{sec:spta} introduces the simulated PTA (SPTA) test configuration used throughout. \S\ref{sec:bases} discretizes the framework into five sky-decomposition bases (pixel, spherical harmonic, square-root spherical harmonic, radiometer, and principal-map). \S\ref{sec:results} covers the angular response of a PTA, various injections of anisotropy, a full Fisher analysis, and anisotropic sensitivity forecasting. We discuss these results and conclude in \S\ref{sec:conclusion}.

Throughout this paper, we adopt the following notational conventions. Subscripts denote discrete indices or labels (pulsar IDs, matrix elements, pixel indices), and parentheses denote continuous arguments (frequency, sky direction, time). Calligraphic symbols ($\mathcal{R}, \mathcal{M}$) denote basis-agnostic objects on the sphere, while their plain counterparts ($R$, $M$) are the discretized versions in a particular basis. The continuous sky direction is $\hat{\Omega}$, while the direction of pixel $k$ in the pixel basis is $\hat{\Omega}_k$. We work throughout in the weak signal regime, with only the Earth term in the pairwise response.

\section{Methods}\label{sec:methods}

Many of the methods used for modeling anisotropy with pulsar timing arrays have been detailed in literature 
\cite{mingarelli_characterizing_2013, taylor_searching_2013, gair_mapping_2014,
ali-haimoud_fisher_2020, ali-haimoud_insights_2021, pol_forecasting_2022,
gersbach_mapping_2025, moursy_anisotropy_2026}, and here we compile the various methods into a single consistent notation and publicly available code. Much of this section is therefore a review, reproduced so that the different approaches and bases are on a common footing. Throughout, we index pulsars by $I,J$, sky directions by $\hat{\Omega}$, and gravitational wave frequency by $f$. 

\subsection{Cross-Correlation Framework}

The optimal cross-correlation estimator for a pulsar pair ($I$,$J$) \cite{allen_detecting_1999, anholm_optimal_2009, chamberlin_time-domain_2015} is
\begin{equation}\label{eqn:cross_corr}
\hat{\rho}_{IJ}=\frac{r_{I}^{T}P_{I}^{-1}\hat{S}_{IJ}P_{J}^{-1}r_{J}}{\mathrm{tr}\left[P_{I}^{-1}\hat{S}_{IJ}P_{J}^{-1}\hat{S}_{JI}\right]},
\end{equation}
where $r_I$ is the timing residual vector for pulsar I, whose elements are the differences between the observed pulse times of arrival and those predicted by the pulsar's deterministic timing model. Here $P_I$ is its noise covariance matrix, and $\hat{S}_{IJ}$ is a template cross-power matrix set by the assumed GWB spectral shape, evaluated between the observation times of pulsars $I$ and $J$ (defined explicitly in Eq.~\eqref{eqn:powermatrix_fd}). For an isotropic GWB, the signal matrix, the expected cross-covariance of the residual vectors $S_{IJ} = \langle r_I\ r_J^T \rangle$, is then
\begin{equation}\label{eqn:signal_matrix}
    S_{IJ} = A^2_{\mathrm{GWB}}\chi_{IJ}\hat{S}_{IJ} = A^2_{\mathrm{GWB}}\tilde{S}_{IJ},
\end{equation}
where $\chi_{IJ}$ is the Hellings-Downs coefficient (the generalization to anisotropic skies follows in the next subsection). The variance of each pair defines the pair uncertainty $\sigma_{IJ}$: 
\begin{equation}\label{eqn:trace_variance}
    \frac{1}{\sigma_{IJ}^{2}}=\mathrm{tr}\left[P_{I}^{-1}\hat{S}_{IJ}P_{J}^{-1}\hat{S}_{JI}\right].
\end{equation}
The expectation value of Eq.~\eqref{eqn:cross_corr} is
\begin{equation}
\langle\hat{\rho}_{IJ}\rangle_{\mathrm{iso}}=\frac{\mathrm{tr}\left[P_{I}^{-1}\hat{S}_{IJ}P_{J}^{-1}S_{JI}\right]}{\mathrm{tr}\left[P_{I}^{-1}\hat{S}_{IJ}P_{J}^{-1}\hat{S}_{JI}\right]},
\end{equation}
which simplifies using Eq.~\eqref{eqn:signal_matrix} to
\begin{equation}
    \langle\hat{\rho}_{IJ}\rangle_{\mathrm{iso}} = A^2_{\mathrm{GWB}}\chi_{IJ}.
\end{equation}

\subsection{Anisotropic Signal Model}

The single-direction antenna pattern functions \cite{romano_detection_2017} for a pulsar $I$ with unit position vector $\hat{p}_I$ (pointing from Earth toward the pulsar) and propagation direction $\hat{\Omega}$ (pointing from the source toward Earth) are:
\begin{equation}
    F_I^A(\hat{\Omega})=\frac{\hat{p}_I^a\hat{p}_I^b}{2(1+\hat{\Omega}\cdot\hat{p}_I)}e_{ab}^A(\hat{\Omega}),\quad A\in\{+,\times\},
\end{equation}
where $e_{ab}^A$ is the gravitational wave polarization tensor for $+$ or $\times$ polarization. The gravitational wave strain $h_I$ as measured at Earth for pulsar $I$ is a superposition over all sky directions and polarizations
\begin{equation}\label{eqn:h_I}
  h_{I}(t) = \int df \int d^{2}\hat{\Omega}\,\Big[ h_{+}(f,\hat{\Omega}) F_{I}^{+}(\hat{\Omega})
  + h_{\times}(f,\hat{\Omega}) F_{I}^{\times}(\hat{\Omega}) \Big]\,e^{-2\pi i f t}.
\end{equation}
We take the strain to be a zero-mean Gaussian with power that depends on both gravitational wave frequency $f$ and direction $\hat{\Omega}$
\begin{equation}
    \langle h_A(f,\hat{\Omega})h_{A'}^*(f^{\prime},\hat{\Omega}^{\prime})\rangle\!\!=\!\!\frac{1}{2}S_h(f)P(\hat{\Omega})\delta_{AA'}\delta(f\!\!-\!\!f^{\prime})\delta^2(\hat{\Omega}\!-\!\hat{\Omega}^{\prime}),
\end{equation}
where the polarization indices $A$, $A' \in \{+, \times \}$ and the Kronecker delta $\delta_{AA'}$ encode an unpolarized background, with equal and uncorrelated power in the two polarizations, consistent with the two-polarization strain of Eq.~\eqref{eqn:h_I} and the response of Eq.~\eqref{eqn:RIJ}. Here $S_h(f)$ is the one-sided strain power spectral density (PSD) and $P(\hat{\Omega})$ is the angular power distribution.\footnote{Two distinct quantities share the symbol $P$, the noise covariance matrix $P_I$, which always carries a capital Latin subscript denoting the pulsar index, and the angular power distribution $P(\hat{\Omega})$, which appears as a function of sky direction.} In principle, $P(\hat{\Omega})$ may itself be a function of frequency, since higher frequencies are expected to grow more anisotropic as $C_\ell/C_0 \propto f^{11/3}$ in the GW-driven regime \cite{sato-polito_exploring_2024}. Here we adopt the standard assumption that the spectral and angular dependence separate, so that $P(f,\hat\Omega)\to P(\hat\Omega)$ for all results in this paper. Because every quantity that follows is built frequency by frequency, this costs no generality (the implementation supports the general $P(f, \hat{\Omega})$ directly), and it condenses each result into a single sky map. 

The pairwise timing response function \cite{ali-haimoud_fisher_2020,ali-haimoud_insights_2021} (also called the overlap response matrix \cite{pol_forecasting_2022}) is then
\begin{equation}\label{eqn:RIJ}
    \mathcal{R}_{IJ}(\hat{\Omega})\equiv\frac{3}{2}\left[F_{I}^{+}(\hat{\Omega})F_{J}^{+}(\hat{\Omega})+F_{I}^{\times}(\hat{\Omega})F_{J}^{\times}(\hat{\Omega})\right],
\end{equation}
and the effective sky-weighted overlap reduction function (ORF) that appears in $\langle\hat{\rho}_{IJ}\rangle$ is 
\begin{equation}\label{eqn:gamma_IJ}
    \Gamma_{IJ}=\int\frac{d^2\hat{\Omega}}{4\pi}\mathcal{R}_{IJ}(\hat{\Omega})P(\hat{\Omega}).
\end{equation}
Now for an anisotropic sky, our expectation value of Eq.~\eqref{eqn:cross_corr} is
\begin{equation}\label{eqn:exp_rho}
\langle\hat{\rho}_{IJ}\rangle=A_{\mathrm{GWB}}^{2}\Gamma_{IJ}=A_{\mathrm{GWB}}^{2}\int\frac{d^{2}\hat{\Omega}}{4\pi}\mathcal{R}_{IJ}(\hat{\Omega})P(\hat{\Omega}).
\end{equation}
For an isotropic sky ($P = 1$), $\Gamma_{IJ}$ reduces to the Hellings-Downs \cite{hellings_upper_1983} coefficient
\begin{equation}
    \chi_{IJ}(\theta)=\frac{3}{2}\left[\frac{1}{3}+\frac{1-\cos\theta}{2}\left(\ln\frac{1-\cos\theta}{2}-\frac{1}{6}\right)\right].
\end{equation}

\subsection{Translating to the Fourier Domain}

If pulsar noise is stationary, the time domain noise covariance matrix $P_I$ has a Toeplitz structure, meaning each descending diagonal is constant and corresponds to the autocorrelation at the relevant time lag \cite{Crisostomi:2025vue}. In the Fourier (gravitational wave frequency) basis that diagonalizes a Toeplitz matrix, $P_I$ becomes a diagonal matrix of power spectral densities
\begin{equation}\label{eqn:toeplitz}
\begin{aligned}
  (P_I)_{tt'} &= \int_{-\infty}^{+\infty} df\,S_I(f)\,e^{2\pi i f(t-t')}, \\
  \tilde{P}_I(f) &= S_I(f),
\end{aligned}
\end{equation}
where $S_I(f)$ is the one-sided strain noise power spectral density of pulsar $I$. Following Eq.~(71) of Hazboun, Romano, \& Smith (2019) \cite{hazboun_realistic_2019}, it is built from the inverse-noise-weighted transmission function $\mathcal{N}_I^{-1}(f)$, which encodes each pulsar's noise covariance and timing model fit, and the single-pulsar gravitational wave response $\mathcal{R}(f)=1/(12\pi^2f^2)$, as
\begin{equation}
\begin{aligned}
    S_I(f) &= \frac{1}{\mathcal{N}_I^{-1}(f)\mathcal{R}(f)} = \frac{12\pi^2f^2}{\mathcal{N}_I^{-1}(f)} \\
    &\approx12\pi^2 f^2\left[2\sigma_I^2\Delta t + A_I^2 f^{-\gamma_I}\right] + S_h(f),
\end{aligned}
\end{equation}
so that the array's real, heterogeneous noise enters every subsequent expression through $S_I(f)$ \cite{hazboun_realistic_2019}. The approximate form follows from Eq.~(24) of \cite{hazboun_realistic_2019}, where $2\sigma_I^2\Delta t$ is pulsar $I$'s white noise and $A_I^2 f^{-\gamma_I}$ is the intrinsic red noise in pulsar $I$ (Eq.~(23) of \cite{hazboun_realistic_2019}). The last term, $S_h(f)$, is the self-noise of the GWB, which includes the power from both the Earth-Earth and pulsar-pulsar autocorrelations (Eqs.~(36) and~(85) of \cite{hazboun_realistic_2019}), and \texttt{hasasia} includes both by design.  Here $\mathcal{R}(f)$ is the isotropic response of a single pulsar to a gravitational wave, distinct from the angular pairwise response $\mathcal{R}_{IJ}(\hat{\Omega})$ of Eq.~\eqref{eqn:RIJ}. Unlike $\mathcal{R}_{IJ}$, it keeps the pulsar term through the pulsar-pulsar autocorrelation (Eqs.~(36) and~(50) of \cite{hazboun_realistic_2019}). Retaining it in $\mathcal{R}_{IJ}$ would only matter for two pulsars close enough in space that a gravitational wave passes both with nearly the same phase \cite{mingarelli_characterizing_2013}. Their distances from Earth would need to agree within about a gravitational wavelength, since otherwise the pulsar-pulsar term averages away across the band \cite{allen_hellings_2023}. Their angular separation on the sky would also need to be within 1 degree for two pulsars at $D\sim1$ kpc \cite{mingarelli_effect_2014}. The pulsar terms of a pair this close would correlate along with their Earth terms, changing the overlap reduction function by tens of percent or more \cite{mingarelli_effect_2014}. The GWB template cross-power matrix $\hat{S}_{IJ}$, shown in Eqs.~\eqref{eqn:cross_corr}-\eqref{eqn:signal_matrix}, now has its frequency content set by the assumed spectral shape $\tilde{S}_h(f)$
\begin{equation}\label{eqn:powermatrix_fd}
    (\hat{S}_{IJ})_{tt^{\prime}}=\int df\tilde{S}_{h}(f)e^{2\pi if(t-t^{\prime})},
\end{equation}
where $t$ runs over the observation times of pulsar $I$ and $t^{\prime}$ over those of pulsar $J$, so the pair dependence of $\hat{S}_{IJ}$ enters only through the two pulsars' sampling, and $S_h(f) = A^2_{\mathrm{GWB}}\tilde{S}_{h}(f)$.\\

If we now substitute Eqs.~\eqref{eqn:toeplitz}-\eqref{eqn:powermatrix_fd} into our Eq.~\eqref{eqn:trace_variance}, the resulting matrices become pointwise products in frequency, and the time-domain trace reduces to an integral over frequency, scaled by the overlapping observation time $T_{IJ}$ of the pulsar pair ($I$, $J$)
\begin{equation}\label{eqn:variance_fd}
    \begin{aligned}\frac{1}{\sigma_{IJ}^2}&=\mathrm{tr}\left[P_{I}^{-1}\hat{S}_{IJ}P_{J}^{-1}\hat{S}_{JI}\right]\\&=T_{IJ}\int_{-\infty}^{+\infty}df\frac{\tilde{S}_{h}(f)}{S_{I}(f)}\cdot\frac{\tilde{S}_{h}(f)}{S_{J}(f)}\\&=2T_{IJ}\int_{0}^{\infty}df\frac{\tilde{S}_{h}^{2}(f)}{S_{I}(f)S_{J}(f)},\end{aligned}
\end{equation}
where the last line folds the integral to positive frequencies only, picking up a factor of 2. 

Eq.~\eqref{eqn:variance_fd} gives us the per-pair variance in the Fourier domain. It weights each frequency by the inverse noise product of the two pulsars and integrates over the observing band.  We use only this per-pair variance and neglect the full covariance between distinct pulsar pairs, the standard weak-signal approximation appropriate for the sensitivity curve regime considered here \cite{allen_detecting_1999, romano_detection_2017, hazboun_realistic_2019}. This approximation assumes that the power a pair shares is small compared with each pulsar's own noise, $|\Gamma_{IJ}|S_h/(S_IS_J)^{1/2} \ll 1$. Understanding the covariance between distinct pulsar pairs beyond this limit is an area of active research \cite{allen_hellings_2023, gersbach_mapping_2025}.

\subsection{Fourier-Domain SNR}

The matched-filter single-frequency signal to noise ratio (SNR) for the power at a single sky direction $\hat{\Omega}$, treated in isolation, seen by pulsar pair ($I$, $J$) is
\begin{equation}\label{eqn:per_freq_snr}
    \mathrm{SNR}_{IJ}^{2}(f,\hat{\Omega}) = 2T_{IJ}\frac{S_{h}^{2}(f)\mathcal{R}_{IJ}^{2}(\hat{\Omega})P(\hat{\Omega})^{2}}{S_{I}(f)S_{J}(f)},
\end{equation}
which is the per-frequency, directional form of the pair statistic $\langle\hat\rho_{IJ}\rangle^2/\sigma_{IJ}^2$, obtained by combining Eqs.~\eqref{eqn:exp_rho} and~\eqref{eqn:variance_fd} with power restricted to the single direction $\hat{\Omega}$ \cite{allen_detecting_1999, anholm_optimal_2009, hazboun_realistic_2019}. It is the building block of everything that follows.\\

The total pair SNR is then
\begin{equation}\label{eqn:total_pair_snr}
  \mathrm{SNR}^2_{IJ} \;\equiv\; \frac{\langle\hat\rho_{IJ}\rangle^2}{\sigma^2_{IJ}},
\end{equation}
with $\langle\hat\rho_{IJ}\rangle$ and $\sigma^2_{IJ}$ given by Eqs.~\eqref{eqn:exp_rho} and~\eqref{eqn:variance_fd}, respectively. Because the sky integral enters through $\Gamma_{IJ}$ inside $\langle\hat\rho_{IJ}\rangle$ before squaring, the total pair SNR is the square of a sky integral rather than the $P(\hat{\Omega})$-weighted sky average of Eq.~\eqref{eqn:per_freq_snr}. The distinction between these two orderings reappears as the radiometer versus full-Fisher split of \S\ref{subsec:fisher_seff}.
  
\subsection{Directional Fisher Matrix and ${S_{\mathrm{eff}}}$} \label{subsec:fisher_seff}

Stripping the common signal factors $2S_h^2(f)P(\hat{\Omega})^2$ from Eq.~\eqref{eqn:per_freq_snr} and generalizing the single-direction response $R_{IJ}^2(\hat{\Omega})$ to the two-direction product $R_{IJ}(\hat{\Omega})R_{IJ}(\hat{\Omega'})$, then summing over pulsar pairs, gives the per-frequency directional Fisher matrix on the sky, following Ali-Ha\"imoud et al. (2020, 2021) \cite{ali-haimoud_fisher_2020,ali-haimoud_insights_2021} and Pol et al. (2022) \cite{pol_forecasting_2022}
\begin{equation}\label{eqn:fisher_continuous}
  \mathcal{M}(\hat\Omega,\hat\Omega';f) = \sum_{I<J} \frac{T_{IJ}}{T_{\mathrm{obs}}}\,
  \frac{\mathcal{R}_{IJ}(\hat\Omega)\,\mathcal{R}_{IJ}(\hat\Omega')}{S_I(f)\,S_J(f)},
\end{equation}
where each pair is weighted by its duty cycle $T_{IJ}/T_{\mathrm{obs}}$ relative to the reference observing time $T_{\mathrm{obs}}$. The diagonal $\mathcal{M}(\hat\Omega,\hat\Omega;f)$ captures the raw information content per direction, and the off-diagonal piece encodes the geometric correlations that the quadrupolar antenna pattern induces between sky locations. From this, two effective sky sensitivities arise. Each is an effective strain-noise power spectral density with units of strain$^2$/Hz, and both are properties of the array alone, independent of any particular signal. The gravitational wave strain is dimensionless, so those units are inverse hertz, matching the strain PSD $S_h(f)$, and the background enters the detection statistics of \S\ref{subsec:total_snr} only through the dimensionless ratio $S_h/S_{\mathrm{eff}}$. Physically, $S_{\mathrm{eff}}$ is to a sky direction what $S_I$ is to a single pulsar, and the characteristic strain built from it is $h_c(f,\hat{\Omega})=\sqrt{f\,S_{\mathrm{eff}}(f,\hat{\Omega})}$. Following Hazboun, Romano, \& Smith (2019) \cite{hazboun_realistic_2019}, we call $S_{\mathrm{eff}}$ and the curves it produces the directional sensitivity curves of this paper's title. The first is the radiometer sensitivity, 
\begin{equation}\label{eqn:seff_rad_continuous}
  \,S_\mathrm{eff}^\mathrm{rad}(f,\hat\Omega)
  = \sqrt{\frac{1}{\mathcal{M}(\hat\Omega,\hat\Omega;f)}}\,,
\end{equation}
which takes the diagonal of $\mathcal{M}$ at a single sky position and then inverts that element, $1/\mathcal{M}(\hat\Omega,\hat\Omega;f)$. Physically, this is the effective sensitivity at a given location assuming there is no signal anywhere else on the sky. \\
The second is the full-Fisher sensitivity, 
\begin{equation}\label{eqn:seff_full_continuous}
  \,S_\mathrm{eff}^\mathrm{full}(f,\hat\Omega)
  = \sqrt{\,\mathcal{M}^{-1}(\hat\Omega,\hat\Omega;f)}\,,
\end{equation}
which inverts $\mathcal{M}$ first, across all sky positions, and then takes the diagonal $\mathcal{M}^{-1}(\hat\Omega,\hat\Omega;f)$. Physically, this is the sensitivity at $\hat{\Omega}$ when the power at every other direction on the sky is left free. 
Because these two orderings do not commute, in general $1/\mathcal{M}(\hat\Omega,\hat\Omega;f)\neq\mathcal{M}^{-1}(\hat\Omega,\hat\Omega;f)$, and they coincide only when $\mathcal{M}$ has no off-diagonal structure to begin with. Neither carries the sky $P(\hat{\Omega})$, and the signal enters through the sky-weighted sensitivity below and the total SNR of \S\ref{subsec:total_snr}.

By the Cauchy-Schwarz inequality on the positive-semidefinite $\mathcal{M}$, at every direction $\hat{\Omega}$, $\mathcal{M}^{-1}(\hat\Omega,\hat\Omega;f) \geq
1/\mathcal{M}(\hat\Omega,\hat\Omega;f)$ \cite{vallisneri_use_2008}. The radiometer's $1/\mathcal{M}(\hat\Omega,\hat\Omega;f)$ is the Cramér-Rao bound for the single-direction point-source hypothesis, and the full-Fisher $\mathcal{M}^{-1}(\hat\Omega,\hat\Omega;f)$ is the Cramér-Rao bound for the free-sky multi-direction hypothesis. The size of the gap between them measures the strength of the off-diagonal couplings in $\mathcal{M}$, i.e., how strongly nearby sky directions share information through general relativity's quadrupolar antenna pattern (see Figure~\ref{fig:fisher_structure}). The radiometer's reliance only on diagonal evaluations of $\mathcal{M}$ also makes it computationally much cheaper than the full Fisher, which requires inverting the full operator. Because $\mathcal{M}$ is built from a finite number of pulsar pairs, it has rank at most $N_{\mathrm{pair}}$ and is not invertible as a continuous operator on the sphere. In practice, $\mathcal{M}^{-1}$ therefore denotes a regularized pseudo-inverse computed in a finite basis whose dimension respects this limit, with the basis choices described in \S\ref{sec:bases} and the regularization scheme detailed in \ref{app:regularization}. Its absolute scale is set by that regularization, so the full-Fisher sensitivity is best read as the marginal uncertainty in distinguishing one direction from the rest of the sky rather than as a detection quantity, while the radiometer of Eq.~\eqref{eqn:seff_rad_continuous}, the sky-weighted curve of Eq.~\eqref{eqn:seff_sky-weighted}, and the principal-map sensitivities of \S\ref{subsec:principal-map} do not rely on the regularization.

A third effective sensitivity follows from contracting $\mathcal{M}$ with the sky on both sides. Unlike the two curves of Eqs.~\eqref{eqn:seff_rad_continuous} and~\eqref{eqn:seff_full_continuous}, it carries the angular power $P(\hat{\Omega})$ and answers a different question: the effective strain-noise for detecting one specified sky, rather than the sensitivity toward a single direction. Weighting $\mathcal{M}$ by $P(\hat{\Omega})$ on both sides and integrating over the sphere gives
\begin{equation}\label{eqn:seff_sky-weighted}
    S_{\rm eff}(f) = \left[\iint\frac{d^2\hat\Omega\,d^2\hat\Omega'}{(4\pi)^2}\,P(\hat\Omega)\,\mathcal{M}(\hat\Omega,\hat\Omega';f)\,P(\hat\Omega')\right]^{-1/2} = \left[\sum_{I<J}\frac{T_{IJ}}{T_{\rm obs}}\frac{\Gamma_{IJ}^2}{S_I(f) S_J(f)}\right]^{-1/2},
\end{equation}
where the second equality follows from the sky-weighted overlap reduction function $\Gamma_{IJ}$ of Eq.~\eqref{eqn:gamma_IJ}. This is the anisotropic generalization of the isotropic stochastic sensitivity, with the Hellings-Downs coefficient $\chi_{IJ}$ promoted to the sky-weighted $\Gamma_{IJ}$, and it requires no inversion of $\mathcal{M}$. In the isotropic limit, where $P(\hat{\Omega})=1$ and $\Gamma_{IJ}\rightarrow\chi_{IJ}$, Eq.~\eqref{eqn:seff_sky-weighted} reduces to
\begin{equation}
S_{\rm eff}(f)=\left[\int\int\frac{d^{2}\hat{\Omega}d^{2}\hat{\Omega'}}{(4\pi)^{2}}\mathcal{M}(\hat{\Omega},\hat{\Omega}^{\prime};f)\right]^{-1/2}=\left[\sum_{I<J}\frac{T_{IJ}}{T_{\mathrm{obs}}}\frac{\chi_{IJ}^{2}}{S_{I}(f)S_{J}(f)}\right]^{-1/2},
\end{equation}
which is exactly the isotropic stochastic gravitational wave background sensitivity, Eq.~(89) of Hazboun, Romano, \& Smith (2019) \cite{hazboun_realistic_2019}.

\subsection{Total SNR}\label{subsec:total_snr}

The signal enters through the angular power $P(\hat{\Omega})$. Integrating the effective sensitivity of Eq.~\eqref{eqn:seff_sky-weighted} over frequency gives the full-Fisher total detection statistic \cite{romano_detection_2017, ali-haimoud_insights_2021, pol_forecasting_2022}, in the same form as the isotropic total SNR, 
\begin{equation}
    \mathrm{SNR}^2_{\rm full} = 2T_{\rm obs}\int df\left(\frac{S_h(f)}{S_{\rm eff}(f)}\right)^2,
\end{equation}
which, with the sky integrals of Eq.~\eqref{eqn:seff_sky-weighted} written out, reads
\begin{equation}\label{eqn:snr_full}
\mathrm{SNR}^2_\mathrm{full} = 2T_{\mathrm{obs}}\int df\,S_h^2(f)
\iint \frac{d^2\hat\Omega\,d^2\hat\Omega'}{(4\pi)^2}\,
P(\hat\Omega)\,\mathcal{M}(\hat\Omega,\hat\Omega';f)\,P(\hat\Omega'),
\end{equation}
and in the isotropic limit $P(\hat{\Omega})=1$, it reduces exactly to Eq.~(90) of Hazboun, Romano, \& Smith (2019) \cite{hazboun_realistic_2019}. Eq.~\eqref{eqn:snr_full} is the matched-filter signal for the sky $P(\hat{\Omega})$, applying $\mathcal{M}$ forward with no inversion required, and is the frequency-resolved, basis-agnostic counterpart of the pixel-basis form $\hat{P}^T [R^T\Sigma^{-1}R]\hat{P}$ used by Pol, Taylor, \& Romano (2022). Discretization in the pixel basis and absorption of the frequency integral into the per-pair inverse-variance weighting $\Sigma^{-1}$, which is diagonal with entries $1/\sigma_{IJ}^{2}$ in the weak signal limit, recovers Eq.~(21) of \cite{pol_forecasting_2022}. More formally, it is the signal term of the frequentist maximum-likelihood detection statistic given in Eqs.~(7.56)-(7.58) of Romano \& Cornish (2017) \cite{romano_detection_2017}. Because that statistic is built from the maximum-likelihood estimate of the sky rather than the true sky, its expected value separates into the same signal term plus a noise floor set by the variance of that estimate. That floor, the variance of the estimated sky contracted with $\mathcal{M}$, equals the number of sky modes the array constrains. The radiometer total is instead obtained by discarding the off-diagonal structure of $\mathcal{M}$ in the chosen basis and adding the resulting single-mode statistics in quadrature. In the pixel basis, which we use for the results in this paper, the per-pixel detection SNR is 
\begin{equation}\label{eqn:snr_perpix}
  \mathrm{SNR}^2(\hat{\Omega}_k) = 2T_{\mathrm{obs}}\int df\,S_h^2(f)\,\frac{P_k^2\,\mathcal{M}(\hat{\Omega}_k,\hat{\Omega}_k;f)}{N_{\mathrm{pix}}^2},
\end{equation}
where $P_k = P(\hat{\Omega}_k)$. This is the familiar directional SNR, with $S_{\mathrm{eff}}^{\mathrm{rad}}$ as the directional noise since $\mathcal{M}(\hat\Omega_k,\hat\Omega_k;f)=1/(S_{\mathrm{eff}}^{\mathrm{rad}})^2$, and it is the quantity shown in the per-pixel SNR skymaps of \S\ref{sec:results}. The radiometer total is its quadrature sum over the sky, 
\begin{equation}\label{eqn:snr_rad}
    \mathrm{SNR}^2_{\mathrm{rad}} = \sum_k \mathrm{SNR}^2(\hat{\Omega}_k) = 2T_{\mathrm{obs}}\int df\,S_h^2(f)\,\frac{1}{N_{\mathrm{pix}}^2}\sum_k P_k^2\,\mathcal{M}(\hat{\Omega}_k,\hat{\Omega}_k;f),
\end{equation}
and the corresponding forms in the other bases follow the same diagonal restriction (\S\ref{sec:bases}). The full-Fisher total instead applies $\mathcal{M}$ forward and adds the off-diagonal coherence between sky directions, which no single-direction sensitivity curve captures. Unlike the full form, the radiometer total depends on the size of the working basis (here the pixelization), because the single-mode statistics of neighboring modes share information. The two are not related by a single inequality. They coincide exactly when the sky power is confined to one pixel, the hypothesis for which the radiometer is optimal (\S\ref{subsec:fisher_seff}), while for extended skies the coherent form gathers the cross-direction correlations the radiometer discards. The contrast is largest in the isotropic limit, where $\mathrm{SNR}_{\mathrm{full}}$ recovers the standard isotropic detection statistic while the radiometer total falls far below it (reported later in Table~\ref{tab:fiso_snr}).

\subsection{Connection to Map-Making}\label{subsec:map_making}

The Fisher operator $\mathcal{M}$ also gives us the dirty-map and clean-map estimators used throughout the PTA anisotropy literature \cite{thrane_probing_2009, gair_mapping_2014, ali-haimoud_insights_2021, semenzato_bias_2025, grunthal_optimising_2026}. The dirty map in direction $\hat{\Omega}$ projects the per-pair cross-correlations $\hat{\rho}_{IJ}$ onto the sky, weighted by their inverse variances $1/\sigma_{IJ}^2$ and the directional response $\mathcal{R}_{IJ}$,
\begin{equation}\label{eqn:dirty_map}
      X(\hat\Omega) = \sum_{I<J}
      \frac{\mathcal{R}_{IJ}(\hat\Omega)\,\hat\rho_{IJ}}{\sigma_{IJ}^2}.
\end{equation}
The clean map $\hat P_{\mathrm{clean}}(\hat\Omega;f)$ is the minimum-variance estimate of the angular power $P(\hat{\Omega})$, obtained by applying the inverse-Fisher operator to the dirty map,
\begin{equation}\label{eqn:clean_map}
      \hat P_{\mathrm{clean}}(\hat\Omega;f) = \frac{1}{T_{\mathrm{obs}}}\int d^2\hat\Omega'\,
      \mathcal{M}^{-1}(\hat\Omega,\hat\Omega';f)\,X(\hat\Omega'),
\end{equation}
with directional uncertainty
\begin{equation}\label{eqn:clean_sigma}
      \sigma(\hat\Omega;f) =
      \sqrt{\frac{\mathcal{M}^{-1}(\hat\Omega,\hat\Omega;f)}{T_{\mathrm{obs}}}} =
      \frac{S_\mathrm{eff}^\mathrm{full}(f,\hat\Omega)}{\sqrt{T_\mathrm{obs}}}.
\end{equation}
The second equality in Eq.~\eqref{eqn:clean_sigma} shows that the full-Fisher effective sensitivity and the directional clean-map uncertainty are the same object up to a factor of $\sqrt{T_{\mathrm{obs}}}$. The radiometer counterpart is $\sigma^\mathrm{rad}(\hat\Omega;f) = 1/\sqrt{T_{\mathrm{obs}}\mathcal{M}(\hat\Omega,\hat\Omega;f)}
= S_\mathrm{eff}^\mathrm{rad}(f,\hat\Omega)/\sqrt{T_\mathrm{obs}}$, which is the single-direction point-source limit of the same family of estimators \cite{mitra_gravitational_2008, thrane_probing_2009, pol_forecasting_2022}, often used alongside Bayesian spherical-harmonic methods \cite{taylor_searching_2013}. As with $S_\mathrm{eff}^{\mathrm{full}}$, $\mathcal{M}^{-1}$ is computed in a finite basis with appropriate regularization in practice. This inverse-Fisher clean map is the standard tool for deconvolved anisotropy reconstruction: it underlies the \texttt{MAPS} formalism \cite{pol_forecasting_2022} used in the NANOGrav 15-year search \cite{agazie_nanograv_2023-2}, and Moursy et al. (2026) \cite{moursy_anisotropy_2026} adopt $\sqrt{\mathrm{diag}(\mathcal{M}^{-1})}$ directly for their empirical sensitivity curves.

\section{The SPTA: A Simulated IPTA-like PTA}\label{sec:spta}

In this work, we use the same simulated dataset that was created in \cite{baier_sensitivity_2025} to be realistic in its noise heterogeneity and similar in scale and sensitivity to the International Pulsar Timing Array (IPTA) \cite{ipta_dr1}. We refer to it throughout as the simulated PTA (SPTA), and it was built to resemble an IPTA configuration with realistic noise. We briefly review it here.

The distribution of pulsars in the simulated PTA is drawn from an empirical distribution created over the IPTA Data Release 3's (DR3) pulsar sky locations to ensure the sky locations of pulsars in the simulated PTA mimic a realistic population of PTA-quality millisecond pulsars. Similarly, the white noise levels in each pulsar are drawn from an empirical distribution of the $\log_{10}$ white noise values reported in \cite{ng15data} and the red noise levels in each pulsar are drawn from a 2D empirical distribution of red noise hyperparameters from \cite{ng15_detchar} when modeled alongside a common uncorrelated red noise (CURN) so as not to double-count the GWB self-noise and intrinsic pulsar red noise. Lastly, we inject an isotropic power-law GWB that is consistent with the background measured in \cite{3pta_comp}. This background is part of each simulated pulsar's noise covariance matrix, so $S_I$ carries its self-noise (Eq.~(85) of \cite{hazboun_realistic_2019}). We construct this dataset with uneven timespans of the pulsars to mimic a realistic scenario of a continually growing pulsar timing array and we adjust the overall white noise levels until the simulated PTA has a GWB SNR of $\sim7$ at a timespan of $16$ years, which roughly matches projections for the IPTA DR3 SNR \cite{baier_sensitivity_2025}.

The simulated array grows over a total span of $40$ years, and we analyze it at two snapshots along that growth. The first is a $16$-year slice, retaining only those pulsars whose timespans exceed $3$ years within that window. The second is the full $40$-year array. The $16$-year snapshot is our simulated stand-in for the present-day IPTA DR3-like array, while the $40$-year snapshot represents its extension into the future. At the \cite{3pta_comp} background amplitude used to tune the noise, the $16$-year slice ($115$ pulsars) has an isotropic background SNR of $\sim7$, and the full $40$-year array ($157$ pulsars) reaches $\sim32$. On top of this tuned array, the results in \S\ref{sec:results} inject a fixed power-law GWB with $A_{\mathrm{GWB}} = 2.4\times10^{-15}$ and $\gamma = 13/3$. Both the $16$-year and $40$-year snapshots satisfy the weak-signal approximation, as the median over all pulsar pairs for $|\Gamma_{IJ}|S_h/(S_IS_J)^{1/2}$ is below 0.07 at every frequency, including the few nanohertz where the GWB is the loudest \cite{siemens_stochastic_2013}. The $16$-year SPTA is most sensitive at $f\approx8.54$ nHz, and we refer to this frequency throughout the remainder of the manuscript as the peak sensitivity frequency, evaluating the single-frequency skymaps of \S\ref{sec:results} there unless stated otherwise.

\section{Anisotropy Bases}\label{sec:bases}

The Fisher object $\mathcal{M}(\hat\Omega,\hat\Omega';f)$ is basis-agnostic, but in practice the sky power $P(\hat{\Omega})$ has to be parameterized by a finite set of coefficients $\{a_\mu\}$ with a chosen basis $\{B_\mu(\hat{\Omega})\}$. The choice of basis is important as each provides a different parameterization of the angular power distribution and no single basis is uniformly the best. We implement five bases on a common Fisher backbone: four in externally chosen coordinates (pixel, spherical harmonic, square-root spherical harmonic, radiometer) and a fifth defined by the array itself, the principal-map basis. Each of the four externally chosen bases reduces to the Hellings-Downs isotropic limit under appropriate normalization. Table~\ref{tab:bases} summarizes the five bases, their primary use cases, and their drawbacks.

\begin{figure}
    \centering
    \includegraphics[width=0.5\linewidth]{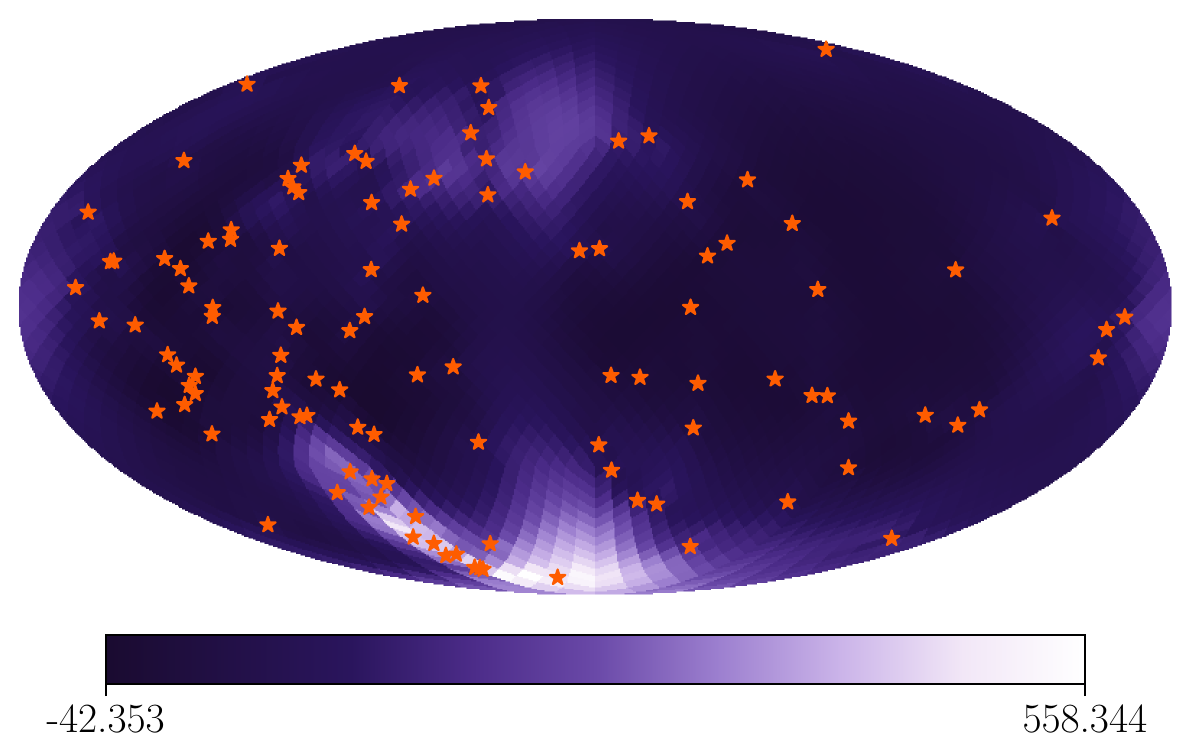}
    \caption{Pixel basis pairwise timing response function, summed over pulsar pairs $\sum_{I<J}{R}_{IJ,k}$ for the 16-year SPTA (115 pulsars). Orange stars denote pulsar locations on the sky. The response peaks towards a dense concentration of pulsars in the southern sky, but it is shaped by the geometry of all pulsar pairs rather than just by local density. }
    \label{fig:RIJ_pixel}
\end{figure}

\subsection{Pixel Basis}

The most direct way of parameterizing the sky power is simply to give its value on a discrete pixel grid,
\begin{equation}
P(\hat{\Omega})=\sum_{k}P_{k}\delta(\hat{\Omega},\hat{\Omega}_{k}),
\end{equation}
where $P_k$ is the power in pixel $k$ on an equal-area HEALPix\footnote{http://healpix.sourceforge.net} grid \cite{gorski_healpix_2005} with $N_{\mathrm{pix}} = 12N_{\mathrm{side}}^2$. We choose $N_{\mathrm{side}}$ using the methods described in Gersbach et al. (2026) \cite{gersbach_mapping_2025}
\begin{equation}\label{eqn:nside}
    N_\mathrm{side}\leq\left[\frac{N_\mathrm{psr}(N_\mathrm{psr}-1)}{24}\right]^{1/2},
\end{equation}
here, we evaluate this upper bound and then round down to the nearest power of 2.

This bound follows the counting argument that the $N_{\mathrm{pair}}$ pulsar-pair cross-correlations carry at most $N_{\mathrm{pair}}$ independent pieces of angular information \cite{ali-haimoud_insights_2021, gersbach_mapping_2025}. Agarwal et al. (2026) caution that tying the resolution to this counting argument can under-resolve an array, and set the recovery multipole instead from the array's own principal maps \cite{agarwal_addressing_2026}, while a multi-resolution pixelization offers a further refinement \cite{moursy_anisotropy_2026}. Since our aim is to demonstrate the framework rather than to characterize any particular array, we adopt this counting-based $N_{\mathrm{side}}$ throughout. The grid is deliberately finer than the angular scales the array can resolve, so the discretization is not the limiting factor.

We can then define the per-pair per-pixel response matrix following the conventions laid out in Pol, Taylor, \& Romano (2022) \cite{pol_forecasting_2022}
\begin{equation}\label{eqn:RIJ_pixel}
    R_{IJ,k}=\frac{\mathcal{R}_{IJ}(\hat{\Omega}_{k})}{N_{\mathrm{pix}}},
\end{equation}
where $R_{IJ,k}$ is the pixelized response and $\hat{\Omega}_k$ is the pixel center direction. Figure~\ref{fig:RIJ_pixel} shows ${R}_{IJ,k}$  for the 16-year SPTA summed over pulsar pairs. The response peaks near a cluster of pulsars in the southern part of the sky, as local density would suggest. Its full shape, however, is set by the combined geometry of all pulsar pairs, since widely separated pairs contribute broad patterns far from either pulsar, so regions with few nearby pulsars still carry the response.

The normalization is chosen here so that an isotropic sky has $P_k=1$ in every pixel, with $\sum_k P_k = N_{\mathrm{pix}}$. Using Eq.~\eqref{eqn:nside} for the 115 pulsar 16-year SPTA gives $N_{\mathrm{side}}\leq 23$, which we round down to the nearest power of two setting $N_{\mathrm{side}}=16$ throughout. 

The corresponding Fisher matrix is the discrete version of Eq.~\eqref{eqn:fisher_continuous}
\begin{equation}\label{eqn:mkk}
    M_{kk^{\prime}}(f)=\mathcal{M}(\hat{\Omega}_k, \hat{\Omega}_{k'};f)
\end{equation}
which is the bridge between the basis-agnostic kernel and the finite-dimensional matrix that the radiometer and full-Fisher estimators actually act on. The pixel-basis effective sensitivities are then the discrete forms of Eqs.~\eqref{eqn:seff_rad_continuous}~and~\eqref{eqn:seff_full_continuous}, 
\begin{equation}\label{eqn:seff_pix_rad}
    S_{\mathrm{eff}}^{\mathrm{rad}}(f,\hat{\Omega}_{k})=\frac{1}{\sqrt{M_{kk}(f)}},
\end{equation}
\begin{equation}\label{eqn:seff_pix_full}
    S_{\mathrm{eff}}^{\mathrm{full}}(f,\hat{\Omega}_{k})=\sqrt{\left[M^{-1}(f)\right]_{kk}},
\end{equation}
the discrete versions of Eqs.~\eqref{eqn:seff_rad_continuous} and~\eqref{eqn:seff_full_continuous}, respectively.

In practice, the pixel basis is the natural setting for radiometer dirty/clean maps and for direct sky imaging,  but a pixel-basis reconstruction can produce negative values of $P_k$ where the array is poorly constrained. This is why the square-root spherical harmonic basis, which enforces $P_k \geq 0$ by construction, is sometimes preferred for parameter inference over the pixel and spherical harmonic bases, neither of which guarantees positivity.

\subsection{Spherical Harmonic Basis}

\begin{figure*}[htb!]
    \centering
    \begin{subfigure}[t]{0.49\textwidth}
        \centering
        \includegraphics[width=\textwidth]{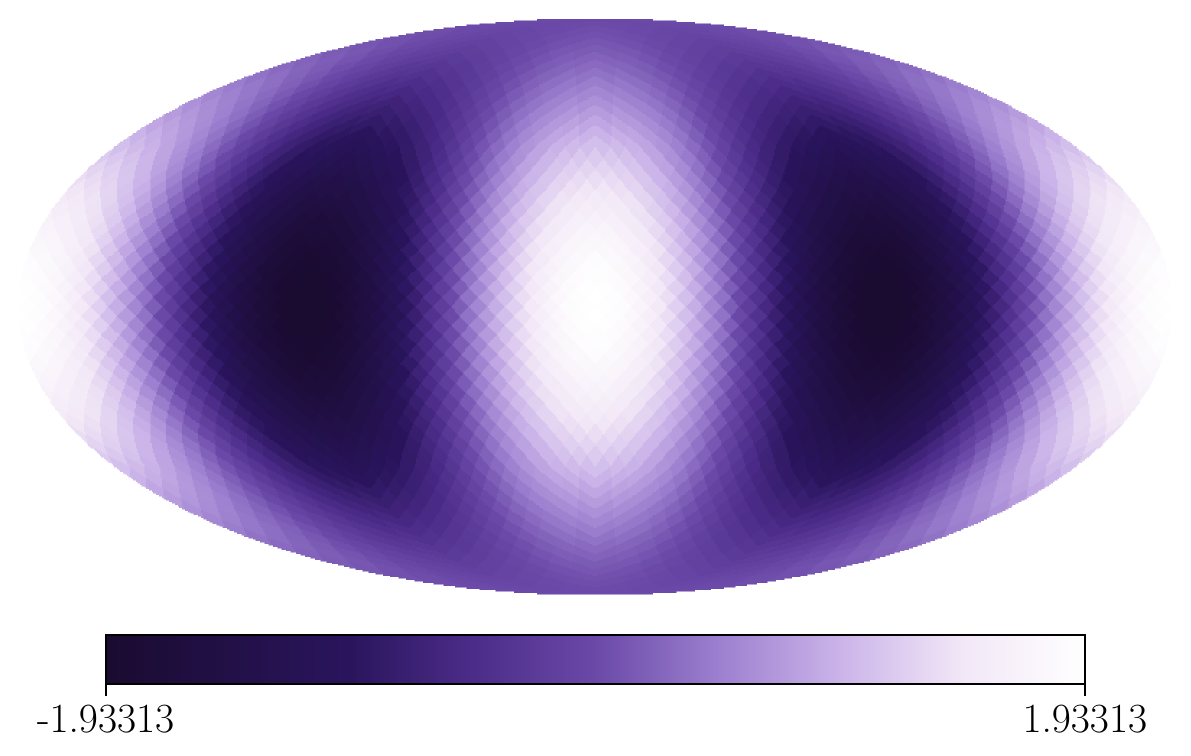}
        \caption{$Y_{\ell m}(\hat{\Omega})$ for $\ell=2, m=2$}
        \label{fig:Ylm_22}
    \end{subfigure}
    \begin{subfigure}[t]{0.49\textwidth}
        \centering
        \includegraphics[width=\textwidth]{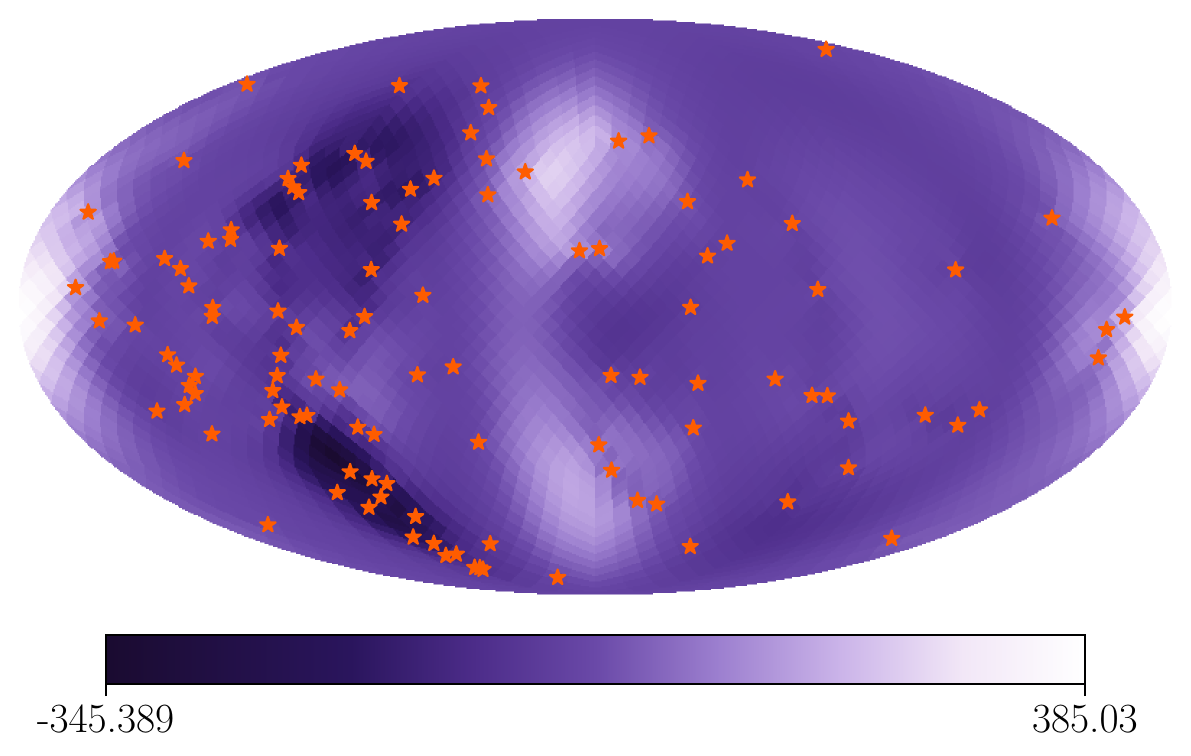} 
        \caption{$\sum_{I<J}{R}_{IJ,\ell=2, m=2}$}
        \label{fig:RIJ_22}
    \end{subfigure}
    \caption{$Y_{2,2}(\hat\Omega)$ (a) and its projection onto the summed response $\sum_{I<J}R_{IJ,\ell m}$ (b).}
    \label{fig:Ylm_RIJ_22}
\end{figure*}

The standard alternative to a per-pixel representation is to expand $P(\hat{\Omega})$ in spherical harmonics in the same style as Mingarelli et al. (2013) \cite{mingarelli_characterizing_2013}
\begin{equation}
    P(\hat{\Omega})=\sum_{\ell=0}^{\ell_{\max}}\sum_{m=-\ell}^{\ell}c_{\ell m}Y_{\ell m}(\hat{\Omega}),
\end{equation}
where the $c_{\ell m}$ are the multipole coefficients and the $Y_{\ell m}(\hat{\Omega})$ are real spherical harmonics normalized so that $Y_{00}=1$, that is, $\sqrt{4\pi}$ times the standard orthonormal harmonics. In this normalization the monopole coefficient $c_{00}=1$ recovers $P=1$ (isotropy), so the monopole mode is the isotropic sky and its effective sensitivity matches the familiar isotropic curve. Pair-coupled multipole projections of $\mathcal{R}_{IJ}(\hat{\Omega})$ give generalized ORFs \cite{mingarelli_characterizing_2013},
\begin{equation}
    \Gamma_{IJ}^{\ell m} = \int\frac{d^2\hat{\Omega}}{4\pi}\mathcal{R}_{IJ}(\hat{\Omega})Y_{\ell,m}(\hat{\Omega}),
\end{equation}

which fall off smoothly with $\ell$, and whose monopole $\Gamma_{IJ}^{00}=\chi_{IJ}$ recovers the Hellings-Downs coefficient. An example of the $\ell=2, m=2$ harmonic can be seen in Figure~\ref{fig:Ylm_RIJ_22} (a), and then the corresponding projection onto ${R}_{IJ,\ell m}$ can be seen in Figure~\ref{fig:Ylm_RIJ_22} (b).

The resulting Fisher matrix in the spherical harmonic basis is
\begin{equation}
    M^{\ell m,\ell^{\prime}m^{\prime}}(f)=\sum_{I<J}\frac{T_{IJ}}{T_{\mathrm{obs}}}\frac{\Gamma_{IJ}^{\ell m}\Gamma_{IJ}^{\ell^{\prime}m^{\prime}}}{S_{I}(f)S_{J}(f)},
\end{equation}
which is the same kernel as Eq.~\eqref{eqn:fisher_continuous} projected onto the spherical harmonic basis through the change-of-basis matrix $U_{k,\ell m} = Y_{\ell m}(\hat{\Omega}) / N_{\mathrm{pix}}$, so that $M^{\ell m,\ell' m'} = U^TMU$ and the monopole element $M^{00,00}$ yields the isotropic effective sensitivity $S_{\mathrm{eff}}=1/\sqrt{M^{00,00}}$. The basis is defined to have $(\ell_{\mathrm{max}} + 1)^2$ total modes, since $\sum_{\ell=0}^{\ell_{\max}}(2\ell+1)=(\ell_{\max}+1)^2$.

Just like with the pixel basis, we can now outline the effective sensitivity in the spherical harmonic basis for both the diagonal radiometer treatment, 
\begin{equation}
    S_{\mathrm{eff}}^{\mathrm{rad,SH}}(f,\ell m)=1/\sqrt{M^{\ell m,\ell m}(f)},
\end{equation}
and the full-Fisher sensitivity,
\begin{equation}
    S_{\mathrm{eff}}^{\mathrm{full,SH}}(f,\ell m)=\sqrt{[M^{-1}(f)]^{\ell m,\ell m}}.
\end{equation}
They can be transformed back into directional skymaps for visualization by recombining the spherical harmonic coefficients on the pixel grid. Similar to the pixel basis, the spherical harmonic basis does not enforce positivity, since a finite $\ell_{\mathrm{max}}$ sum can go negative even for a non-negative sky, as can be seen by looking at the $\ell=2, m=2$ mode projected onto the antenna pattern in Figure~\ref{fig:Ylm_RIJ_22}. A second and more subtle limitation is spectral leakage \cite{semenzato_bias_2025}. Because any practical expansion has to be truncated at a finite $\ell_{\mathrm{max}}$, choosing that cutoff below the array's own angular resolution leaks small-scale angular power down into the multipoles, an effect driven by the way the array's anisotropic response couples different multipoles together, and one that can bias the recovery of any anisotropy. This leakage affects any truncated basis, not the spherical harmonic expansion alone, and is largely avoided by choosing $\ell_{\mathrm{max}}$ at least as large as the array's angular resolution. Because the array's response already suppresses small-scale power on its own, a finite cutoff at that resolution captures what the array can actually measure \cite{agarwal_addressing_2026}.

\subsection{Square-Root Spherical Harmonic Basis}

To enforce $P(\hat{\Omega}) \geq 0$ by construction, one expands $\sqrt{P(\hat{\Omega})}$ rather than $P(\hat{\Omega})$ directly \cite{taylor_bright_2020}. Following the formulation of Pol, Taylor, \& Romano (2022) \cite{pol_forecasting_2022}, we can write the square-root spherical harmonic power distribution as:
\begin{equation}\label{eqn:sq_sphharm_p}
    P(\hat{\Omega})=\left[\sum_{L=0}^{L_{\max}}\sum_{M=-L}^{L}a_{LM}Y_{LM}(\hat{\Omega})\right]^{2}\geq0,
\end{equation}
where the $a_{LM}$ are the square-root spherical harmonic coefficients. This translates to the ordinary multipole coefficients through Clebsch-Gordan products:
\begin{equation}
    c_{\ell m} = \sum_{LM, L'M'} a_{LM}a_{L'M'}\,\beta_{\ell m}^{LM, L'M'},
\end{equation}
with the Gaunt coefficient defined in \cite{konstandin_prospects_2026} as
\begin{equation}
    \beta_{\ell m}^{LM, L'M'} = \sqrt{\frac{(2L + 1) (2L' + 1)}{4\pi(2\ell + 1)}}C_{LM,L'M'}^{\ell m}C_{L0,L'0}^{\ell 0},
\end{equation}
and where $C_{LM,L'M'}^{\ell m}$ are the Clebsch-Gordan coefficients \cite{pol_forecasting_2022, konstandin_prospects_2026}. In Eq.~\eqref{eqn:sq_sphharm_p}, squaring the amplitude field multiplies pairs of harmonics $Y_{LM}Y_{L'M'}$, and the Clebsch-Gordan coefficients expand each product onto power multipoles $\ell$ in the range $|L-L'|\le\ell\le L+L'$. Because the amplitude field is fit only up to degree $L_\mathrm{max}$, the largest accessible power multipole is the top of this range with $L=L'=L_\mathrm{max}$, giving $\ell_\mathrm{max}=L+L'=2L_\mathrm{max}$. The power map is therefore band-limited at $2L_\mathrm{max}$, while the number of independently fit coefficients remains $(L_{\mathrm{max}}+1)^2$. As with the ordinary spherical harmonic basis, this truncation leaks small-scale power into the retained multipoles when $L_{\mathrm{max}}$ falls below the array's angular resolution.

Due to the square-root spherical harmonic basis enforcing positivity, it is quite effective at point-source recovery, and has been used in the NANOGrav 15-year anisotropy search \cite{agazie_nanograv_2023-2} and subsequent work \cite{gersbach_mapping_2025, konstandin_prospects_2026}. In this basis the mapping from the coefficients $a_{LM}$ to $P(\hat{\Omega})$ is quadratic rather than linear. The Fisher matrix for the $a_{LM}$ is therefore no longer a simple linear transformation of the directional Fisher matrix $\mathcal{M}$, as it is in the pixel and spherical harmonic bases where that mapping is linear. Consequently, the direct radiometer and full-Fisher $S_{\mathrm{eff}}$ methods of the previous bases do not carry over unmodified at the $a_{LM}$ level. In practice, this leads to the square-root spherical harmonic basis being used primarily at the inference stage, rather than as a direct sensitivity curve basis. 

\subsection{Radiometer Basis}

The radiometer basis is the simplest we examine here \cite{mitra_gravitational_2008, thrane_probing_2009, agazie_nanograv_2023-2, gersbach_mapping_2025}. It collapses the search to a single pixel at a time, neglecting all off-diagonal Fisher correlations and effectively setting the power to zero in all other pixels. This gives a per-pixel power estimate and uncertainty
\begin{equation}
\hat{P}_{k}^{(\mathrm{rad})}=X_{k}/(T_{\mathrm{obs}}M_{kk}),\quad\sigma_{k}^{(\mathrm{rad})}=1/\sqrt{T_{\mathrm{obs}}M_{kk}},
\end{equation}
where $X_k$ is the dirty map from \S\ref{subsec:map_making} and $M_{kk}$ is the diagonal Fisher information from Eq.~\eqref{eqn:mkk}.

This means the radiometer is optimal when the signal is confined to a single pixel, in which case the off-diagonal couplings carry no information. The radiometer is also computationally inexpensive because it requires no matrix inversion, making it a common companion to Bayesian spherical-harmonic searches \cite{taylor_searching_2013}.

As discussed in \S\ref{subsec:fisher_seff}, the radiometer is the optimal estimator for the single-direction point-source hypothesis, in which the power in every other pixel is held at zero. When the true sky is multi-pixel, that hypothesis is mismatched. By discarding the off-diagonal Fisher components, the radiometer then under-reports the true per-pixel uncertainty by precisely the Cauchy-Schwarz gap between $1/M_{kk}$ and $[M^{-1}]_{kk}$. Throughout this paper we report the radiometer alongside the full-Fisher estimator, which propagates the off-diagonal geometric couplings the radiometer discards. We demonstrate that the gap between the two is a central diagnostic of the array's angular response.

\subsection{Principal-Map Basis}\label{subsec:principal-map}

The four preceding bases all parametrize the sky power in externally chosen coordinates: i.e., a pixel grid, spherical harmonics, their square root, or a single non-zero pixel. A fifth option, introduced in Ali-Ha\"imoud et al. (2020, 2021)\cite{ali-haimoud_fisher_2020,ali-haimoud_insights_2021}, instead lets the array define its own basis. One expands $P(\hat{\Omega})$ in the eigenmodes of the array's Fisher matrix, the principal maps $\hat{m}^{(n)}$ (the $\mathcal{M}_n$ of Ali-Ha\"imoud et al. (2020), written here in lowercase to distinguish them from the Fisher matrix $M$), 
\begin{equation}
    P(\hat{\Omega})=\sum_{n}a_{n}\hat{m}^{(n)}(\hat{\Omega}),
\end{equation}
where $a_n$ corresponds to the amplitude of mode $n$. Following the steps outlined in Ali-Ha\"imoud et al. (2021),  one can work in any of the preceding bases at a fixed frequency and eigendecompose the Fisher matrix,
\begin{equation}
    M_{\mu\nu}(f)\hat{m}_{\nu}^{(n)}(f)=\lambda_{n}(f)\hat{m}_{\mu}^{(n)}(f),\quad\lambda_{n}=1/\Sigma_{n}^{2},
\end{equation}
where $\Sigma_n^2$ is the noise variance of mode $n$. Because $M$ is real and symmetric, the principal maps are orthonormal and statistically uncorrelated,
\begin{equation}
    \hat{m}^{(n)}\cdot M\hat{m}^{(n^{\prime})}=\lambda_{n}\delta_{nn^{\prime}},\quad M=\sum_{n}\lambda_{n}\hat{m}^{(n)}\otimes\hat{m}^{(n)},
\end{equation}
the spectral decomposition of the Fisher matrix (Eqs.~(65)-(66) of \cite{ali-haimoud_insights_2021}). The modes are ordered by increasing noise $\Sigma_n$, equivalently $\lambda_1 \geq \lambda_2 \geq ...$ since $\lambda_n = 1/\Sigma_n^2$, running from most to least detectable.

The reason this basis is a natural fit for the radiometer versus full-Fisher language of this paper is that the Fisher matrix is diagonal in the principal-map basis. The off-diagonal couplings that opened the Cauchy-Schwarz gap between $S_{\mathrm{eff}}^{\mathrm{rad}}$ and $S_{\mathrm{eff}}^{\mathrm{full}}$ in the pixel and spherical harmonic bases are, by construction, absent here. So the radiometer and full-Fisher effective sensitivities coincide,
\begin{equation}
    S_{\mathrm{eff}}^{\mathrm{rad}}(f,n)=S_{\mathrm{eff}}^{\mathrm{full}}(f,n)=\frac{1}{\sqrt{\lambda_{n}(f)}}=\Sigma_{n}(f),
\end{equation}
giving a single effective sensitivity per principal map. Because each $\Sigma_n$ is a single eigenvalue, it does not move with the truncation threshold, making this the one basis in which the full-Fisher sensitivity carries no regularization choice. This collapse of the two estimators onto one curve is unique to the principal-map basis. In every other basis, the off-diagonal couplings force $S_{\mathrm{eff}}^{\mathrm{rad}}<S_{\mathrm{eff}}^{\mathrm{full}}$, whereas here each mode is constrained independently by its own Fisher eigenvalue $\lambda_n$.

The total $\mathrm{SNR}^2$ is then a plain sum over modes
\begin{equation}
    \mathrm{SNR}^2 = \sum_n \frac{a_n^2}{\Sigma_n^2},
\end{equation}
(corresponding to Eq.~(69) of \cite{ali-haimoud_insights_2021}). Two features follow from the modes being uncorrelated. First, each amplitude $a_n$ has its own noise $\Sigma_n$ that does not depend on the others, so an upper limit on any single mode is a standalone number. This means that adding or dropping other modes will leave it unchanged, unlike in the pixel or spherical harmonic bases where fitting more parameters inflates the uncertainty on each. Second, the eigenvalues fall off steeply: in Ali-Ha\"imoud et al. (2021), they find the mode noise $\Sigma_n$ grows roughly exponentially beyond the first $\sim 20$ modes, so the array measures only a handful of modes well and the rest are noise-dominated. Keeping just those leading modes will then capture most of what the array can constrain, making the principal-map basis a natural setting for detection and for dimensionality reduction.

However, the principal-map basis is limited by its interpretability. A pixel or a $Y_{\ell m}$ is the same object for every array, so two PTAs can compare results directly. This cannot be done with principal maps which are built from a single array's pulsar positions and noise, and the basis changes with frequency. So the modes of two PTAs are different shapes on the sky with no common reference, and no single $\hat{m}^{(n)}$ corresponds to a recognizable structure. The principal-map basis is therefore better suited to detection and forecasting, which need only scalar statistics like the mode-summed SNR above, than to the characterization of the sky, which needs a map or power spectrum in coordinates other arrays can reproduce, which is the goal of this paper. 

\begin{table}[htb]
\caption{\label{tab:bases} Summary of the five bases for representing anisotropy discussed in this paper, each implemented on a common Fisher formalism. We present the parameterization of sky power $P(\hat{\Omega})$, the primary use case, and the drawbacks for each basis.}
\footnotesize
\setlength{\tabcolsep}{4pt}
\begin{tabularx}{\linewidth}{@{}l l X X@{}}
\br
Basis & $P(\hat\Omega)$ & Use case & Drawback\\
\mr
Pixel & $\displaystyle\sum_k P_k\,\delta(\hat\Omega,\hat\Omega_k)$ &
Direct sky imaging, radiometer dirty/clean maps, intuitive directional sensitivity &
Does not ensure positivity, over-parametrized relative to the $\sim \ell^2_{\mathrm{eff}}$ constrained modes, near singular $M$ needing regularization\\[2pt]
Spherical harmonic & $\displaystyle\sum_{\ell m} c_{\ell m} Y_{\ell m}$ &
Diffuse anisotropic GWB, reports $C_\ell/C_0$, compact at low $\ell_\mathrm{max}$ &
Does not ensure positivity, and leaks small-scale power
into the retained multipoles when $\ell_{\mathrm{max}}$ is set below the array's angular resolution\\[2pt]
Square-root \\Spherical harmonic & $\displaystyle\Big[\sum_{LM} a_{LM} Y_{LM}\Big]^2$ & Ensures positivity by construction, useful for point-source / hotspot recovery &
Quadratic $a_{LM}\to P$ map, no direct $S_\mathrm{eff}$ at the $a_{LM}$ level, less transparent inference\\[2pt]
Radiometer & single $P_k$ (others $=0$) &
Very fast all-sky scan, optimal for point-sources, companion to Bayesian spherical harmonic searches &
Discards off-diagonal couplings from $M$, sub-optimal for multi-pixel skies\\[2pt]
Principal-map & $\displaystyle\sum_n a_n\,\hat m^{(n)}$ &
Detection-statistic ranking, dimensionality reduction as the radiometer $=$ full-Fisher by construction, and the only basis needing no regularization, since each $\Sigma_n$ is a standalone eigenvalue &
Array- and frequency-specific, not comparable across PTAs, lossy sky reconstruction\\
\br
\end{tabularx}
\end{table}

\subsection{Basis Normalization}

The angular power $P_k$ can be normalized in more than one way; the conventions differ only by a simultaneous rescaling of $P_k$ and $R_{IJ,k}$, while the product $\sum_k R_{IJ,k}P_k=\Gamma_{IJ}$ is convention-independent. \texttt{hasasia} supports both the literature convention of Pol, Taylor, \& Romano (2022) and Konstandin et al. (2026) $P_k=1$, $\sum_k P_k=N_\mathrm{pix}$ and a more probabilistic way of parameterizing the sky ($P_k=1/N_\mathrm{pix}$, $\sum_k P_k=1$).

\section{Results}\label{sec:results}

Throughout this section, we set $P_k = 1$ to denote a sky-isotropic background, so that the results characterize the intrinsic angular sensitivity of the SPTA itself, before any astrophysical injection is layered on top. These maps describe the array's angular response to different regions of the sky, set by the pulsar locations. The SPTA serves as a worked example; our aim is to demonstrate the framework, not characterize this particular array.

\begin{figure*}
    \centering
    \begin{subfigure}[t]{0.32\textwidth}
        \centering
        \includegraphics[width=\textwidth]{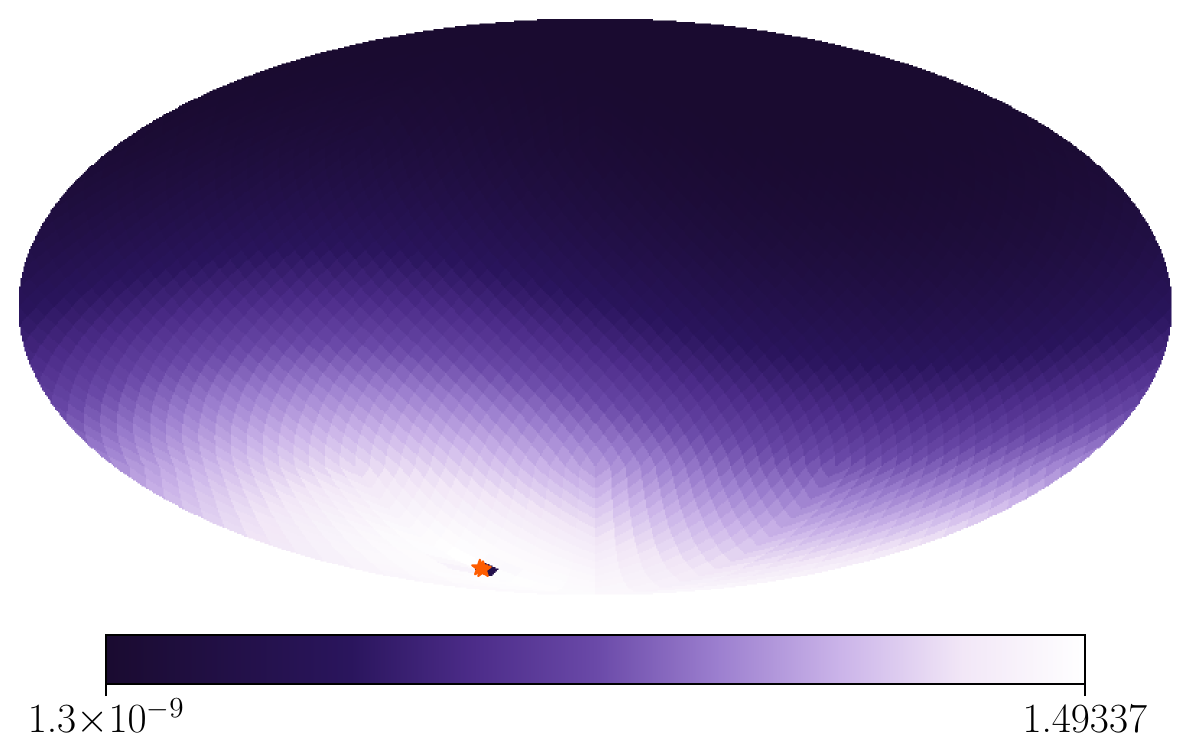}
        \caption{Closest ($1^\circ$)}
        \label{fig:pair_closest}
    \end{subfigure}
    \begin{subfigure}[t]{0.32\textwidth}
        \centering
        \includegraphics[width=\textwidth]{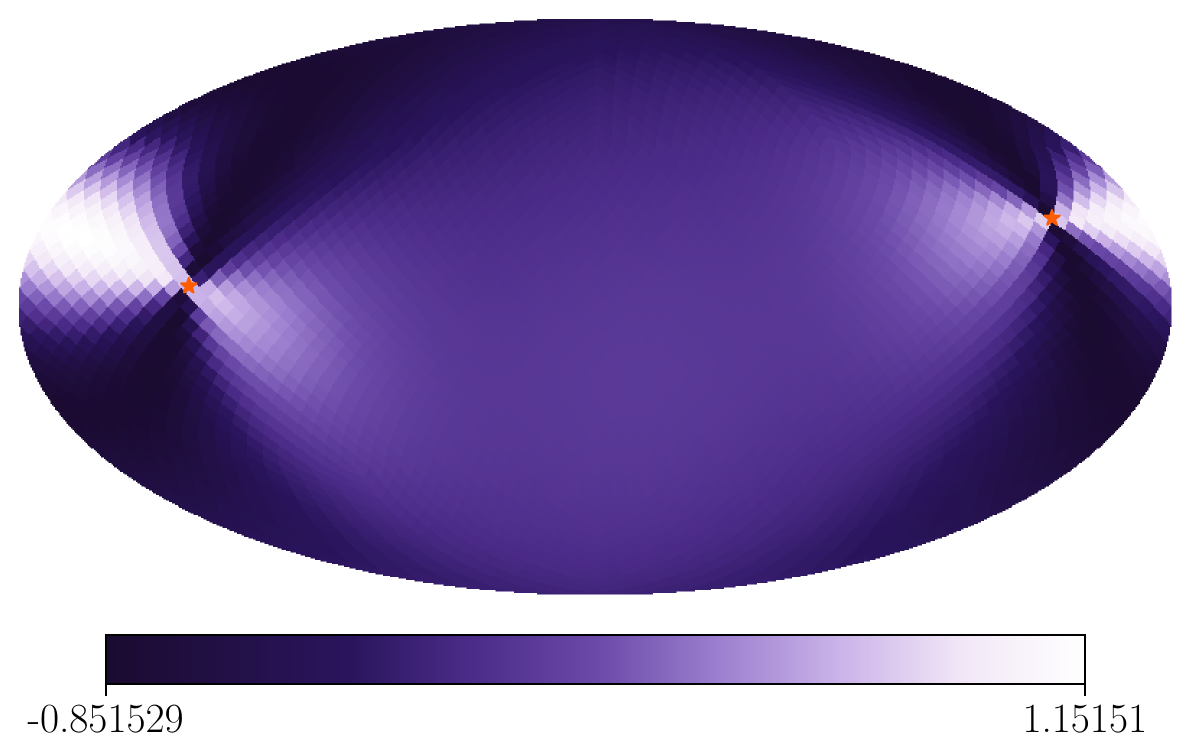} 
        \caption{Median ($82^\circ$)}
        \label{fig:pair_median}
    \end{subfigure}
    \begin{subfigure}[t]{0.32\textwidth}
        \centering
        \includegraphics[width=\textwidth]{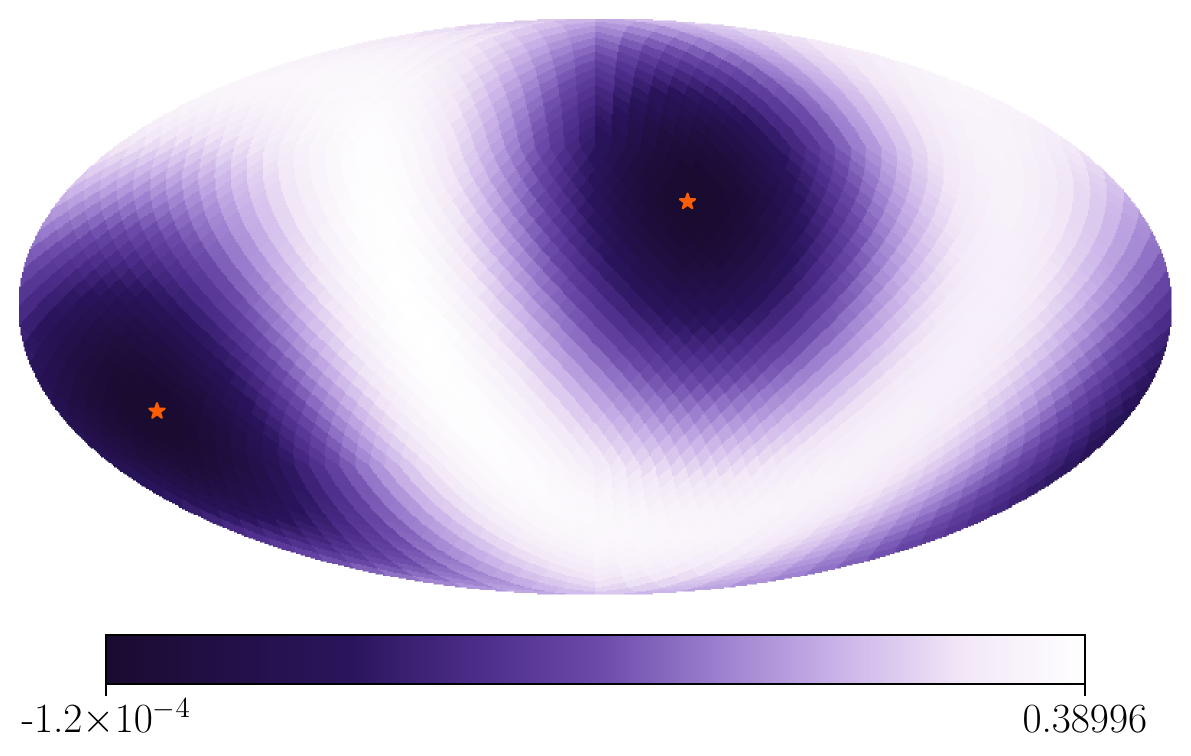}
        \caption{Farthest ($178^\circ$)}
        \label{fig:pair_farthest}
    \end{subfigure}
    \caption{Individual pulsar pair responses $\mathcal{R}_{IJ}$ for representative pairs spanning the full range of angular separations: (a) closest pair ($\sim 1^\circ$), (b) median separation ($\sim 82^\circ$), (c) farthest separation ($\sim 178^\circ$). The single-pair response varies systematically with angular separation, from a concentrated bright region for the closest pair to the clearest two-lobed quadrupolar pattern at the median separation, and a broad banded pattern for the farthest.}
    \label{fig:pair_three_figs}
\end{figure*}

\subsection{Angular Response}

Figure~\ref{fig:RIJ_pixel} shows the summed pixel-basis response $\sum_{I<J}R_{IJ,k}$ for the 16-year SPTA. The response is concentrated near the pulsar clusters, and the regions of the sky farthest away from any pulsars are correspondingly the worst served by the array. Individual pulsar pairs make this more concrete. Figure~\ref{fig:pair_three_figs} highlights representative pulsar pairs that correspond to the closest, median, and farthest separation among the array. The single-pair response changes character with separation: for the closest pair, it collapses to a single bright region near the two near-coincident pulsars (panel a), at median separation it shows the clearest two-lobed quadrupolar pattern near each pulsar (panel b), and for the farthest separation, it spreads into a broad, banded pattern across the sky (panel c). 

\begin{figure*}[htb!]
    \centering
    \begin{subfigure}[t]{0.49\textwidth}
        \centering
        \includegraphics[width=\textwidth]{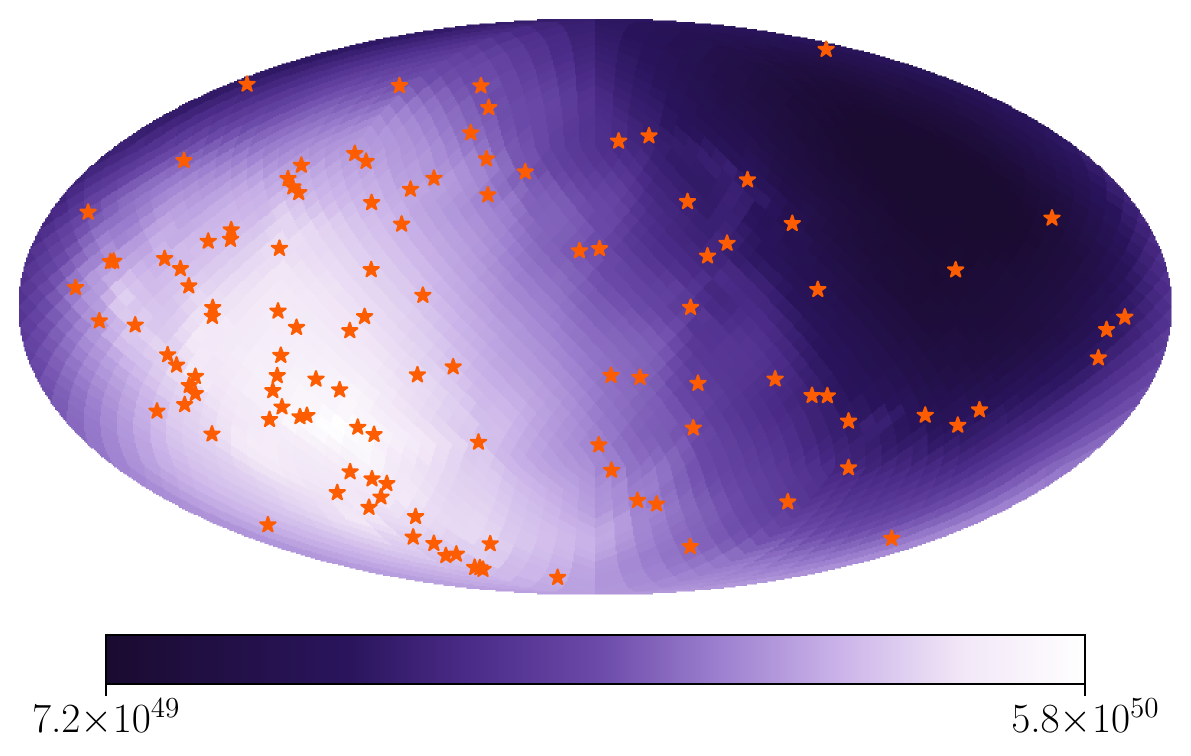}
        \caption{Pixel basis}
        \label{fig:mkk_pixel_skymap}
    \end{subfigure}
    \begin{subfigure}[t]{0.49\textwidth}
        \centering
        \includegraphics[width=\textwidth]{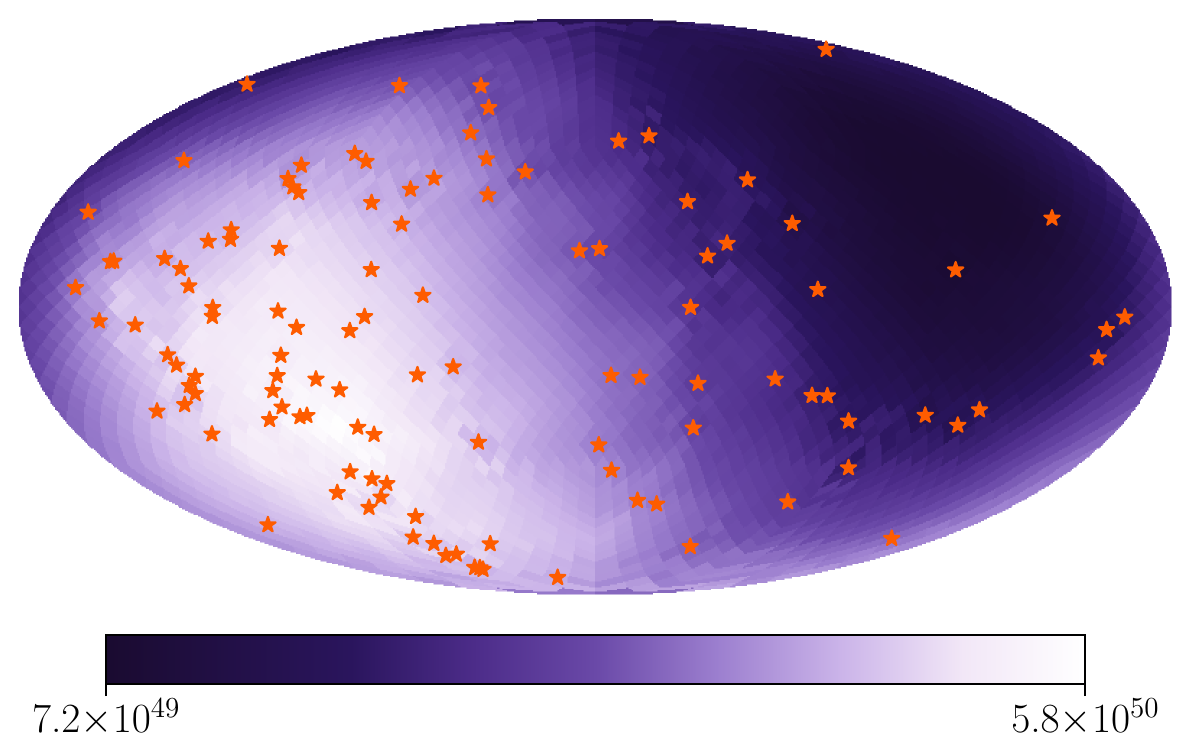} 
        \caption{Spherical Harmonic reconstruction ($\ell_{\mathrm{max}} = 47$)}
        \label{fig:mkk_sh_skymap}
    \end{subfigure}
    \caption{(a) Pixel basis representation of the diagonal Fisher information matrix $M_{kk}$ (shown on a log color scale) at the peak sensitivity frequency ($f \approx 8.54$ nHz) for the 16-year SPTA. (b) The same diagonal Fisher information reconstructed from its spherical-harmonic transform $M(\hat\Omega)=\sum_{\ell=0}^{\ell_{\max}}\sum_{m=-\ell}^{\ell}M_{\ell m}\,Y_{\ell m}(\hat\Omega)$, truncated at $\ell_{\mathrm{max}} = 47 = 3 N_\mathrm{side}-1$, the band limit of the $N_\mathrm{side}=16$ grid. Both panels share the same color scale; at this $\ell_{\mathrm{max}}$ the reconstruction is a faithful re-expression of the pixel map, and the two agree to better than 1\%, a consistency check that they project the same underlying $\mathcal{M}$. The Fisher information varies by nearly an order of magnitude across the sky, with the highest values near the dense clusters of pulsars.}
    \label{fig:mkk_skymaps}
\end{figure*}

The diagonal Fisher information $M_{kk}(f)$ at the peak sensitivity frequency in the pixel-basis is shown in Figure~\ref{fig:mkk_skymaps} (a) and the corresponding spherical harmonic reconstruction at $\ell_{\mathrm{max}} = 47$ is shown in (b). Here $\ell_\mathrm{max}=47=3N_\mathrm{side}-1$ is the band limit of the $N_\mathrm{side}=16$ grid, chosen so that the reconstruction faithfully re-expresses the pixel skymap. This is distinct from the physical $\ell_{\mathrm{max}} \approx 6$ adopted for anisotropy elsewhere, and is used here only to test the discretization. The dynamic range across the sky is nearly an order of magnitude, driven largely by the cluster of pulsars on the left side of the HEALPix Mollweide projection plot.  

\begin{figure}
    \centering
    \includegraphics[width=0.5\linewidth]{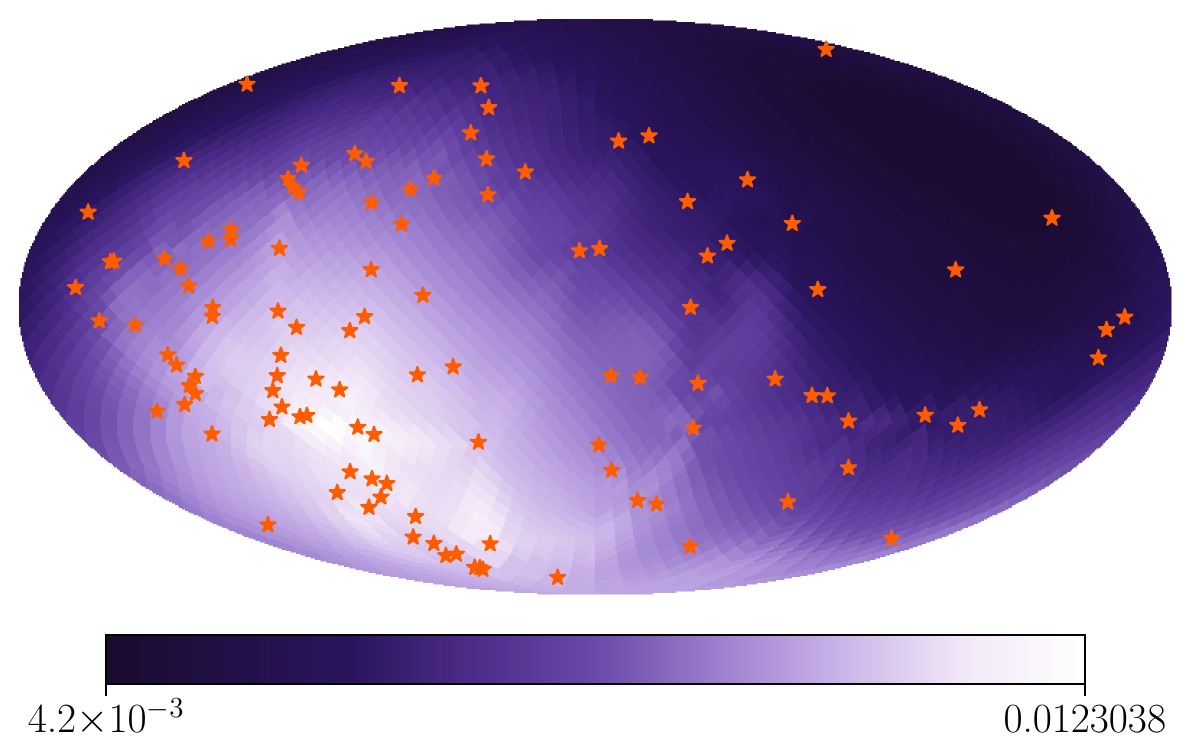}
    \caption{Frequency-integrated per-pixel radiometer detection SNR (Eq.~\eqref{eqn:snr_perpix}) for an isotropically injected GWB ($A_{\mathrm{GWB}} = 2.4 \times 10^{-15}$, $\gamma = 13/3$, $P_k = 1$) for the 16-year SPTA. The map traces the array's $\sqrt{M_{kk}}$ (square root of Figure~\ref{fig:mkk_pixel_skymap}) angular response, peaking in pulsar-dense regions. As a consistency check, the full-Fisher total detection SNR (Eq.~\eqref{eqn:snr_full}) in this isotropic limit is $10.22$, matching the standard isotropic \texttt{hasasia} value to better than $0.01\%$. The radiometer total SNR is far smaller, $\approx 0.44$, because an isotropic sky is maximally multi-pixel, so the diagonal estimator discards the coherent cross-correlations that carry nearly all of the diffuse signal. This coherence gap between the totals is distinct from the Cauchy-Schwarz ordering of the directional sensitivities (Figure~\ref{fig:seff_ratio}), which compares the same two estimators direction by direction.}
    \label{fig:snr_iso_perpixel}
\end{figure}

As a consistency check that the $P_k=1$ limit reduces to the isotropic case, Figure~\ref{fig:snr_iso_perpixel} shows the per-pixel detection SNR for an isotropic injection; the full-Fisher total SNR recovers the standard \texttt{hasasia} value, while summing the per-pixel SNRs in quadrature instead recovers the (smaller) radiometer total (Table~\ref{tab:fiso_snr}).

\begin{figure*}
\centering
\begin{subfigure}[t]{0.48\textwidth}
\centering
\includegraphics[width=\textwidth]{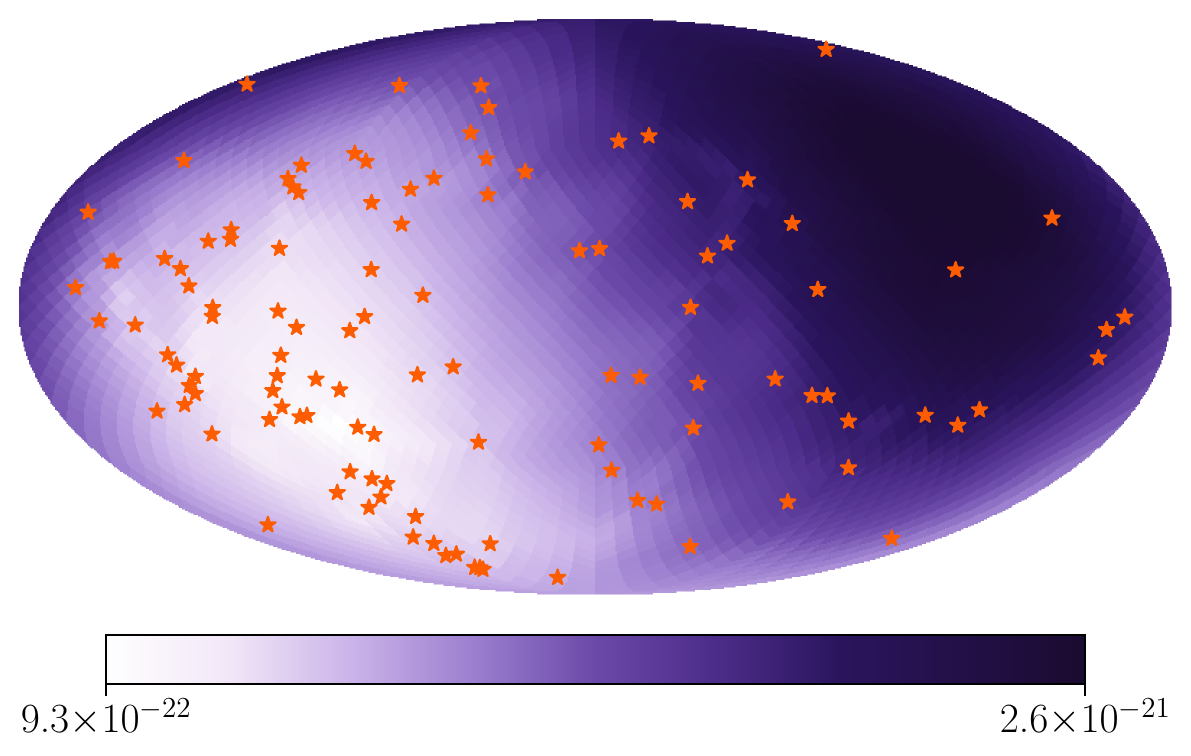}
\caption{Radiometer $S_{\mathrm{eff}}^\mathrm{rad}$}
\label{fig:seff_radiometer}
\end{subfigure}\hfill
\begin{subfigure}[t]{0.48\textwidth}
\centering
\includegraphics[width=\textwidth]{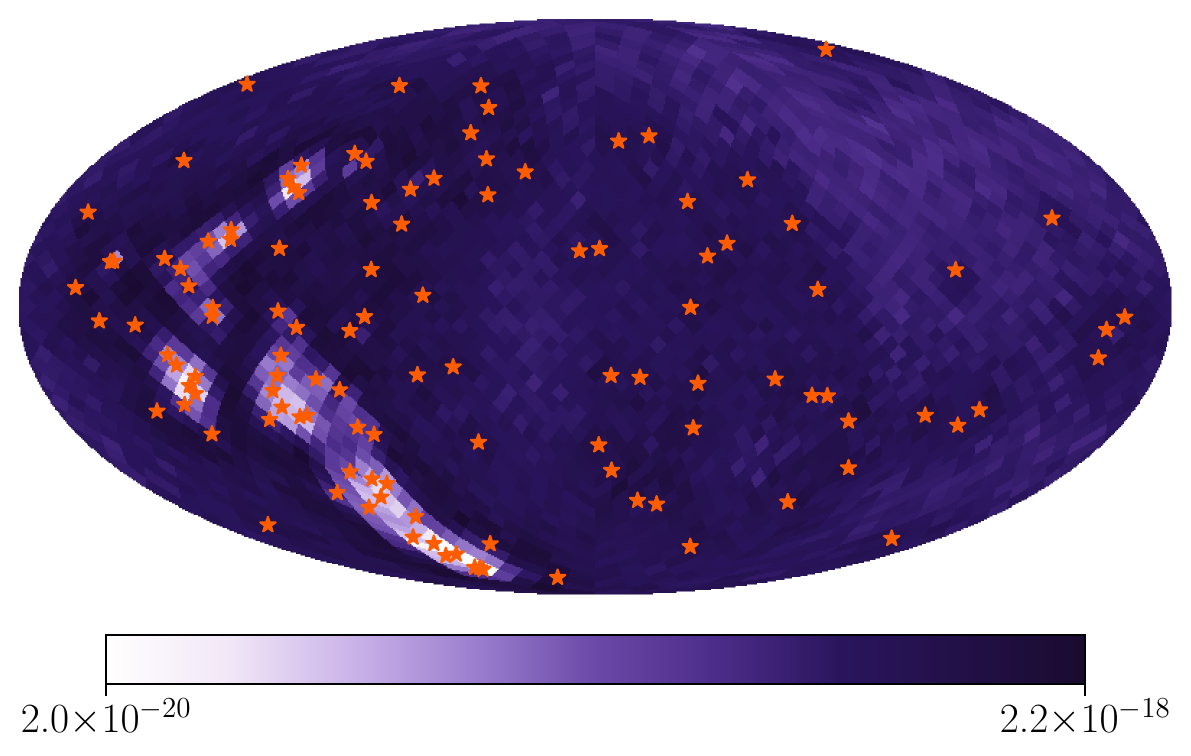}
\caption{Full Fisher $S_{\mathrm{eff}}^\mathrm{full}$}
\label{fig:seff_full}
\end{subfigure}
\vspace{0.5em}
\begin{subfigure}[t]{0.48\textwidth}
\centering
\includegraphics[width=\textwidth]{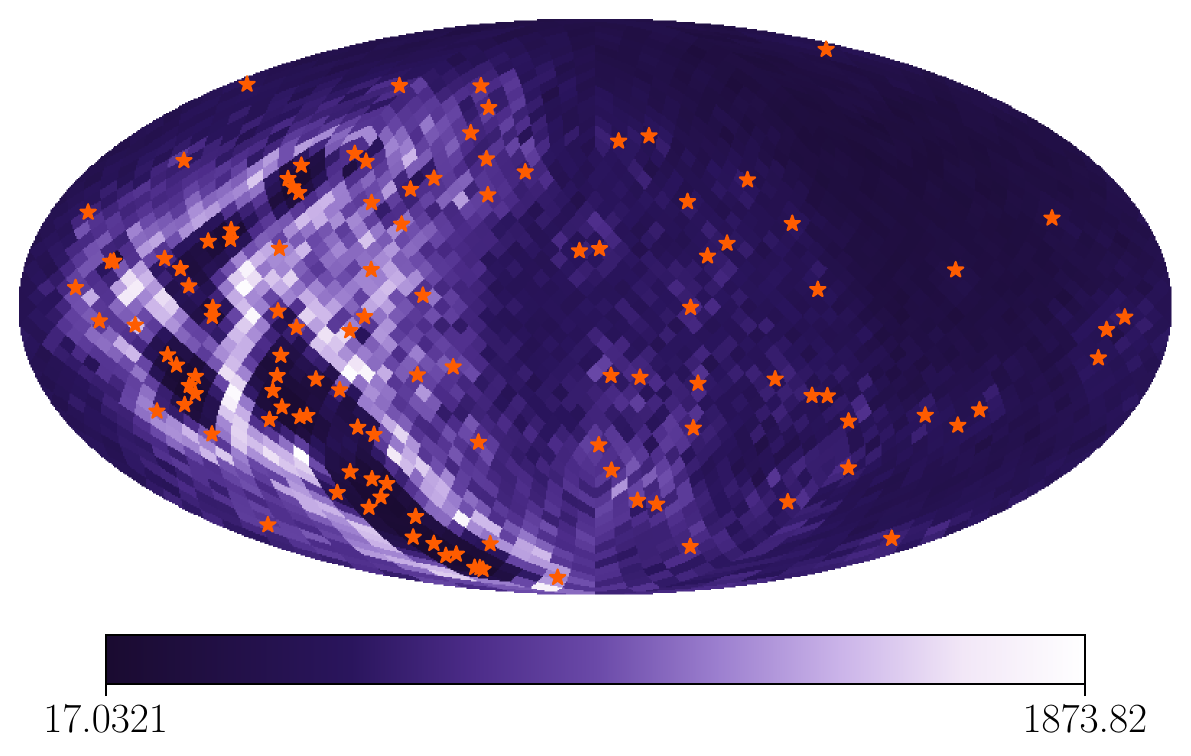}
\caption{Ratio $S_{\mathrm{eff}}^\mathrm{full}$/$S_{\mathrm{eff}}^\mathrm{rad}$}
\label{fig:seff_ratio_panel}
\end{subfigure}
\caption{Per-pixel effective sensitivity $S_{\mathrm{eff}}$ at the peak sensitivity frequency for the 16-year SPTA. (a) shows the diagonal radiometer $S_{\mathrm{eff}}^\mathrm{rad}$ from Eq.~\eqref{eqn:seff_pix_rad}. (b) shows the full-Fisher $S_{\mathrm{eff}}^\mathrm{full}$ from Eq.~\eqref{eqn:seff_pix_full}. (c) shows the ratio of the full-Fisher to the radiometer, which reflects the over-parameterization of the pixel basis relative to the array's $\sim \ell^2_{\mathrm{eff}}$ independent modes.}
\label{fig:seff_ratio}
\end{figure*}

\subsection{Fisher Matrix Analysis}

\begin{figure*}
    \centering
    \includegraphics[width=\linewidth]{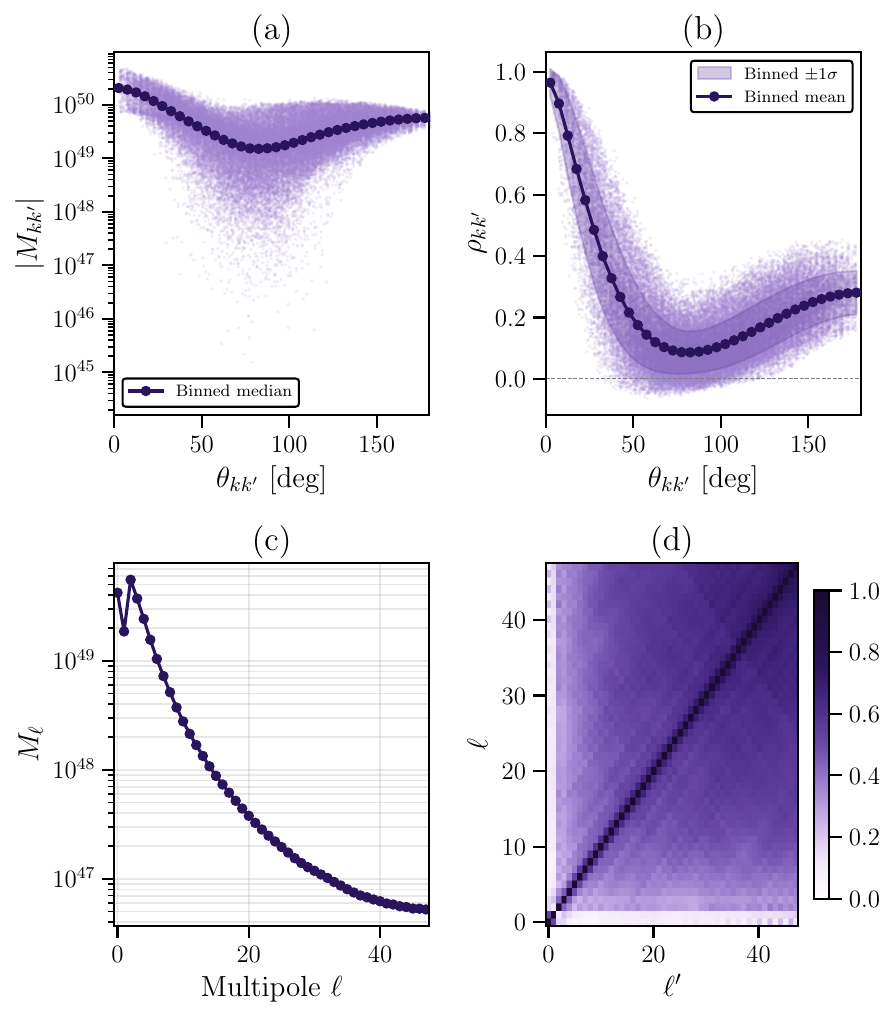}
    \caption{Structure of the Fisher information matrix $M$ for the 16-year SPTA at the peak sensitivity frequency, in the pixel basis (top) and spherical harmonic basis (bottom). (a) $|M_{kk'}|$ versus pixel-pair angular separation $\theta_{kk'}$, with a binned median. (b) Normalized pixel correlation $\rho_{kk'} = M_{kk'}(M_{kk}M_{k'k'})^{-1/2}$, with a binned mean. (c) Per-multipole Fisher information $M_\ell = \sum_m M^{\ell m,\ell m}$ versus $\ell$. (d) Normalized cross-block coupling $\rho_{\ell\ell'} = \|M^{\ell\ell'}\|_F\,(\|M^{\ell\ell}\|_F\|M^{\ell'\ell'}\|_F)^{-1/2}$ between $(\ell,\ell')$ Fisher blocks ($\rho_{\ell\ell}=1$ by construction). Panels (c) and (d) parallel the harmonic-space bias-matrix analysis of Agarwal et al.\ (2026) \cite{agarwal_addressing_2026}, here evaluated for an array with realistic heterogeneous per-pulsar noise rather than the uniform-noise idealization adopted there.}
    \label{fig:fisher_structure}
\end{figure*}

Here we look more closely at the structure of the Fisher information matrix, comparing the radiometer and full-Fisher treatments side by side. In Figure~\ref{fig:seff_ratio}, we show the skymaps for Eqs.~\eqref{eqn:seff_pix_rad} and~\eqref{eqn:seff_pix_full}. As a reminder, in the radiometer treatment of the Fisher matrix, we take the diagonal of the Fisher matrix and find the effective sensitivity in each pixel assuming zeroes in the others. In the full-Fisher we account for the cross-correlations between sky directions, leaving the power in every other pixel free. By Cauchy-Schwarz, the full-Fisher sensitivity is worse (larger $S_{\mathrm{eff}}$) than the radiometer at every pixel, with equality only where a pixel decouples from the rest of the sky. In panel (c), we show the ratio $S_{\mathrm{eff}}^{\mathrm{full}} /S_{\mathrm{eff}}^{\mathrm{rad}}$, which reflects the over-parameterization of the pixel basis relative to the array's $\sim \ell_{\mathrm{eff}}^2$ independently constrained sky modes, and the absolute value of the ratio depends on the singular value decomposition (SVD) truncation. 

This near-singularity is one of conditioning rather than rank. Our choice $N_{\mathrm{side}} = 16$ keeps the pixel count below the pair count ($N_{\mathrm{pix}}=3072 < N_{\mathrm{pair}}=6555$ for the $16$-year array), so $M$ is not rank-deficient through the pair count \cite{ali-haimoud_insights_2021}. Its eigenvalue spectrum instead decays steeply, since the array constrains only the $\sim\ell_{\mathrm{eff}}^2$ independent sky modes set by the quadrupolar antenna response (Figure~\ref{fig:fisher_structure}), leaving the remaining directions measured with rapidly vanishing sensitivity. The full-Fisher inversions therefore require regularization, whose scheme and effect on the reported quantities we detail in \ref{app:regularization}.

A full analysis of the Fisher structure is shown in Figure~\ref{fig:fisher_structure}, which unpacks the inter-pixel and inter-multipole structure of $M$. The left column conveys the absolute scale of the Fisher information, with the pixel basis spanning more than five orders of magnitude in $|M_{kk'}|$, and the per-multipole information $M_\ell$ spanning $\sim3$ orders of magnitude across the multipole range. The right column normalizes each onto a correlation that exposes the inter-mode structure directly. In the pixel basis, the off-diagonal elements are strong at small angular separation ($\theta < 30^\circ$) and decay through zero by $\theta \sim 60^\circ$, while the spherical-harmonic block-correlation metric makes the inter-multipole couplings visible directly. This broad off-diagonal support is precisely what makes the full-Fisher and radiometer estimators differ at the per-pixel level: the array does not measure each pixel (or each multipole) independently. To see why, consider a sky with power in only two pixels, $P_1$ and $P_2$, and everywhere else $P=0$. The normalized coupling is $\rho_{12}=M_{12}/(M_{11}M_{22})^{1/2}$, which is the quantity plotted in panel b. The radiometer estimates $P_1$ by taking the dirty map $X_1$ (Eq.~\eqref{eqn:dirty_map}) and dividing it by $T_{\mathrm{obs}}M_{11}$, treating pixel 2 as if its power were zero. Its estimate of the power in pixel 1 is biased by the actual power in pixel 2, $\langle\hat{P}_1\rangle=P_1 + (M_{12}/M_{11})P_2$. The full-Fisher estimate fits both pixels at the same time and removes that bias, but its uncertainty is larger by a factor of $S_{\mathrm{eff}}^{\mathrm{full}}/S_{\mathrm{eff}}^{\mathrm{rad}}=(1-\rho_{12}^2)^{-1/2}$. Adjacent pixels, at the small-separation end of panel b, have a median coupling of $\rho_{12}=0.974$, so even for them this uncertainty factor is greater than four. More broadly, $\rho_{kk'}$ falls to half its peak by an angular separation $\theta_{1/2} \approx 28^\circ$, setting an effective angular resolution $\ell_{\mathrm{eff}}=\pi/\theta_{1/2}\approx6$. The array therefore constrains only $\sim\ell_{\mathrm{eff}}^2$ independent sky modes, far fewer than the $N_{\mathrm{pix}}=3072$ pixel parameters, which is the over-parameterization that drives the large full-Fisher/radiometer ratio in Figure~\ref{fig:seff_ratio}. The same finite-resolution structure appears in harmonic space: the steep falloff of $M_{\ell}$ (panel c) and the broad off-diagonal coupling of the $\rho_{\ell\ell'}$ blocks (panel d) reflect the leakage and mode-suppression that Agarwal et al. (2026) \cite{agarwal_addressing_2026} show bias the recovered angular power spectrum. They quantify this for an idealized array with uniform per-pulsar noise; the structure we find here is the realistic-noise counterpart, with the heterogeneous noise of the SPTA imprinting additional, configuration-dependent coupling beyond the geometric response alone. 

\subsection{Directional Sensitivity Curves}

\begin{figure}
  \centering
  \includegraphics[width=0.6\linewidth]{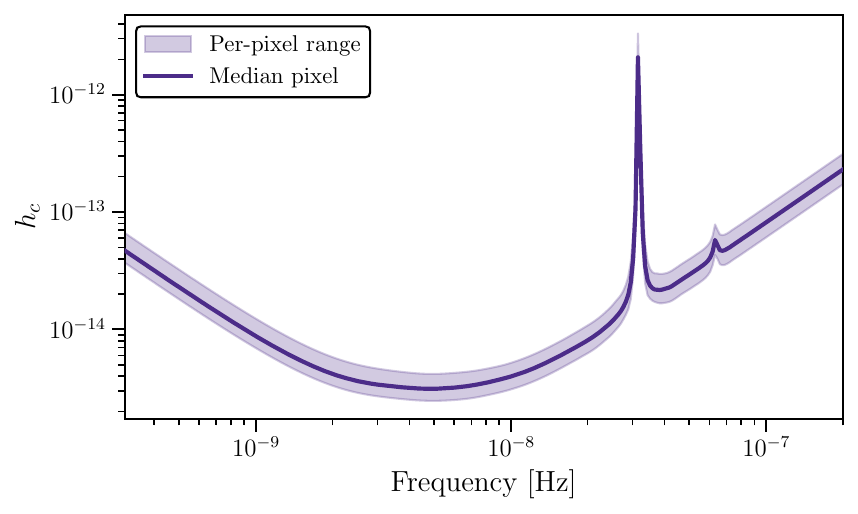}
  \caption{Per-pixel characteristic-strain sensitivity curves $h_c(f,\hat\Omega)$ in the pixel basis for the 16-year SPTA. The shaded band spans the best to worst sky pixels (ranked at the peak sensitivity frequency); the solid curve is the median pixel. Each curve is plotted across the full frequency band.}
  \label{fig:per_pixel_hc}
\end{figure}

\begin{figure}
  \centering
  \includegraphics[width=0.6\linewidth]{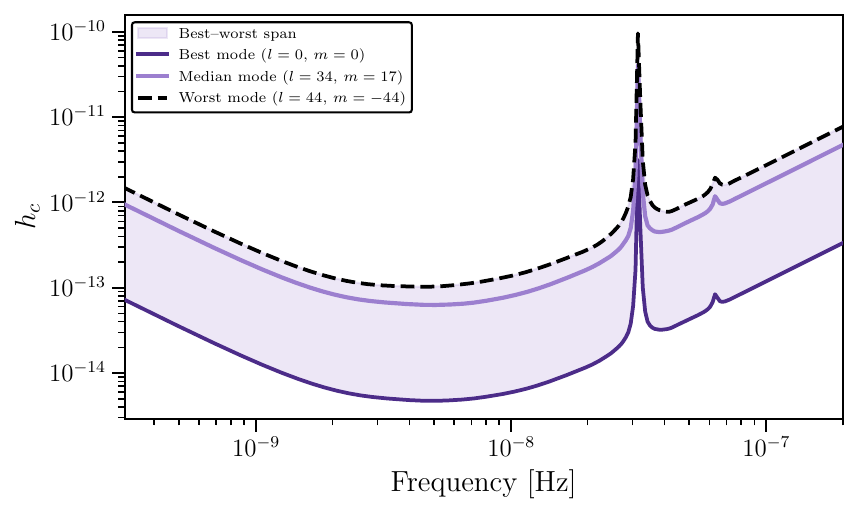}
  \caption{Per-mode characteristic-strain sensitivity curves in the spherical harmonic basis for the 16-year SPTA, showing the best, median (solid), and worst (dashed) $(\ell,m)$ modes ranked at the peak sensitivity frequency. The most sensitive is the monopole ($\ell=0,m=0$), which recovers the isotropic case. The reconstruction uses $\ell_\mathrm{max}=47$; the $\ell=44$ worst mode is simply the least-sensitive mode at this frequency, not a truncation limit.}
  \label{fig:per_mode_hc}
\end{figure}

Beyond skymaps at a single frequency, the directional Fisher framework yields a full characteristic strain sensitivity curve $h_c(f,\hat{\Omega})$. The characteristic strain can be computed from $S_{\mathrm{eff}}$ in the usual way \cite{hazboun_realistic_2019}:
\begin{equation}\label{eqn:hc_rad}
     h_c^{\mathrm{rad}}(f,\hat\Omega_k) = \sqrt{f\,S_{\mathrm{eff}}^{\mathrm{rad}}(f,\hat\Omega_k)}
     = \left(\frac{f}{\sqrt{M_{kk}(f)}}\right)^{1/2},
\end{equation}
with the full-Fisher curve $h_c^{\mathrm{full}}$ given by the same expression with $S_{\mathrm{eff}}^{\mathrm{rad}}\to S_{\mathrm{eff}}^{\mathrm{full}}$ (Eq.~\eqref{eqn:seff_pix_full}).

Figures~\ref{fig:per_pixel_hc} and~\ref{fig:per_mode_hc} show these curves in the pixel basis (per direction) and the spherical harmonic basis (per mode), respectively. To highlight the spread, we rank the pixels and modes by their sensitivity at the peak sensitivity frequency and plot the best, median, and worst, together with the band they span; each curve is then plotted across the full frequency range. The most sensitive mode is the monopole ($\ell=0,m=0$), which recovers the isotropic case. The dynamic range between the best and worst directions is set by the array's angular response: directions near pulsar clusters reach lower $h_c$, while regions of the sky with sparser pulsar coverage are markedly less sensitive. 

\begin{figure}
    \centering
    \includegraphics[width=0.6\linewidth]{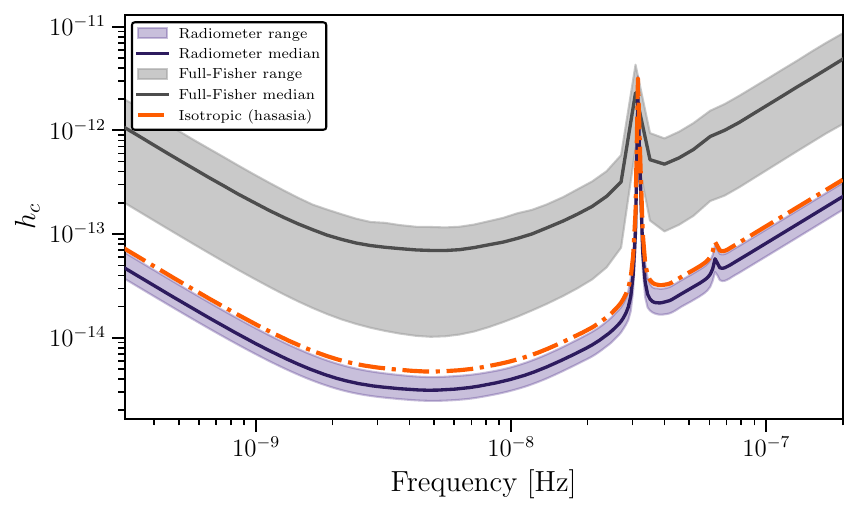}
    \caption{The isotropic sensitivity curve bracketed by the two directional estimators for the 16-year SPTA. The radiometer per-pixel envelope (purple) lies entirely below the isotropic \texttt{hasasia} curve (dash-dot), while the full-Fisher envelope (gray) lies entirely above it. Both envelopes have the median value plotted over the range.}
    \label{fig:radio_iso_full_perdirection}
\end{figure}

The directional curves invite direct comparison with the familiar isotropic sensitivity curve, and the two estimators relate to it in opposite ways. The radiometer and full-Fisher estimators bracket the isotropic curve from opposite sides, as shown in Figure~\ref{fig:radio_iso_full_perdirection}. The radiometer envelope (point-source-optimal) lies entirely below the isotropic curve, while the full-Fisher envelope (which marginalizes each direction over the rest of the sky) lies entirely above it. The outer bracketing is the Cauchy-Schwarz ordering $S_{\mathrm{eff}}^{\mathrm{rad}}\le S_{\mathrm{eff}}^{\mathrm{full}}$ of \S\ref{subsec:fisher_seff}. The full-Fisher's advantage in the total detection statistic (Figure~\ref{fig:snr_iso_perpixel}) is a separate coherence effect: there it wins by retaining the cross-direction correlations, while here, direction by direction, the same off-diagonal couplings make it the more conservative sensitivity.

\subsection{Response to Injected Anisotropy}

We inject a known anisotropic sky $P(\hat{\Omega})$ and quantify the array's directional sensitivity and detection SNR to it, given the realistic per-pulsar noise. Both follow forward from the Fisher operator $\mathcal{M}$ and measure how well the array would detect a specified sky. Here we apply $\mathcal{M}$ forward to a given sky, so the maps and SNR totals are the expected response for that sky, not angular power estimated from noise realizations. Due to this, we make no direct comparisons between the input sky and the estimated one, but instead provide the machinery for this as the inverse-Fisher clean map in \S\ref{subsec:map_making}.

\subsubsection{von Mises-Fisher (vMF) Hotspot Injection Model}

To probe the framework's response to anisotropic signals, we now drop the $P_k = 1$ assumption and inject hotspots on top of an isotropic floor. A localized excess of power in a known direction is a standard anisotropic test signal in the PTA Fisher literature \cite{ali-haimoud_fisher_2020, ali-haimoud_insights_2021}, physically motivated by discrete local supermassive black hole binaries, whether a single loud source or several close together on the sky \cite{mingarelli_local_2017}. Those works parameterize the hotspot differently. We use the von Mises-Fisher (vMF) distribution \cite{fisher_dispersion_1953, mardia_directional_2010}, the spherical analog of a Gaussian:

\begin{equation}
    P(\hat{\Omega}_k)=N_{\mathrm{pix}}\left[f_{\mathrm{iso}}\frac{1}{N_{\mathrm{pix}}}+(1-f_{\mathrm{iso}})\sum_iw_i\mathrm{vMF}_i(\hat{\Omega}_k;\mu_i,\kappa_i)\right],
\end{equation}

where $f_{\mathrm{iso}}$ is the isotropic-floor fraction, $w_i$ are the per-hotspot weights, $\mu_i$ is the localization, and $\kappa_i$ is the concentration of the $i$-th hotspot, with the weighted mixture normalized as a whole to unit sum over pixels, ranging from a sharp, point-source-like hotspot at large $\kappa$ to a broad, diffuse one at small $\kappa$. We now discuss three cases: a single hotspot, two symmetric hotspots, and two asymmetric hotspots. 

\subsubsection{Single Hotspot}

\begin{figure*}
    \centering
    \begin{subfigure}[t]{0.49\textwidth}
        \centering
        \includegraphics[width=\textwidth]{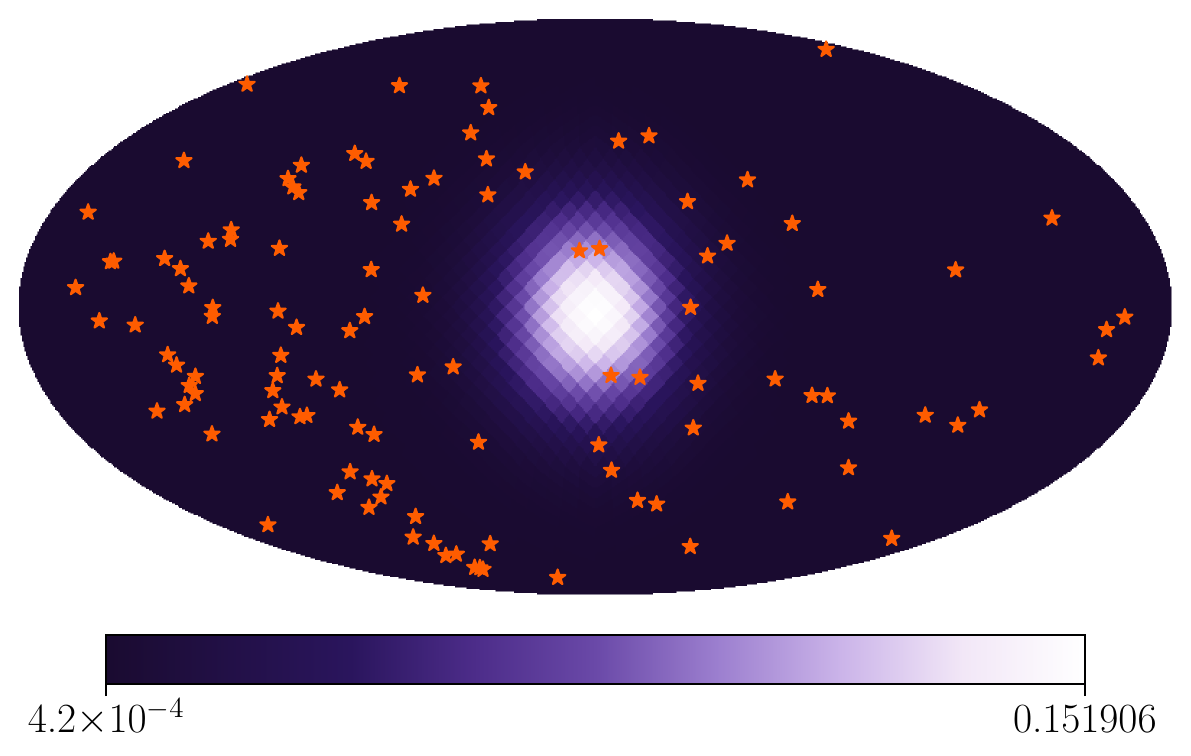}
        \caption{SNR Single Hotspot}
        \label{fig:snr_perfreq}
    \end{subfigure}
    \begin{subfigure}[t]{0.49\textwidth}
        \centering      \includegraphics[width=\textwidth]{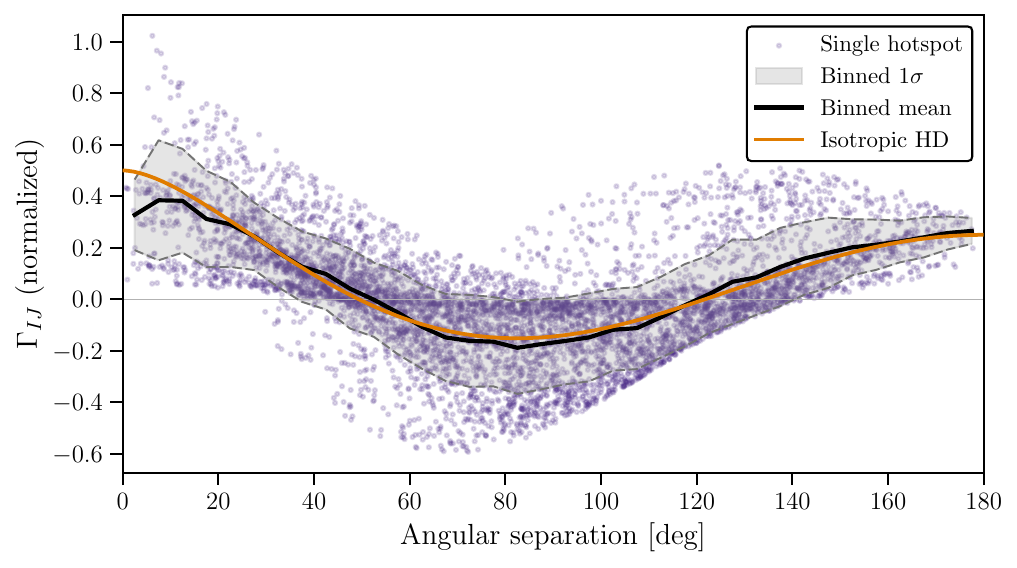} 
        \caption{Per-pair ORF}
        \label{fig:snr_cumulative}
    \end{subfigure}
    \caption{Expected response of the array to a single von Mises-Fisher hotspot ($\kappa=10$, $f_{\mathrm{iso}}=0.10$) injected at $(\mathrm{RA}, \mathrm{Dec})=(180^\circ, 0^\circ)$ on top of an injected GWB with ($A_{\mathrm{GWB}}=2.4\times10^{-15}$, $\gamma=13/3$) for the 16-year SPTA. (a) Frequency-integrated per-pixel radiometer detection SNR (Eq.~\eqref{eqn:snr_perpix}). The expected SNR is modulated by the array's angular response, not by the hotspot alone. (b) The corresponding overlap-reduction function: per-pair effective ORF $\Gamma_{IJ}$ for the hotspot sky (purple scatter), with the binned mean (black) and $1\sigma$ band (gray), compared to the isotropic Hellings-Downs curve (orange). The total detection SNR for this configuration is $1.34$ for the radiometer (Eq.~\eqref{eqn:snr_rad}) and $16.83$ for the full-Fisher (Eq.~\eqref{eqn:snr_full}).}
    \label{fig:snr_hd_single_hotspot}
\end{figure*}

For a single vMF hotspot, the per-pixel quantities (the injected power $P_k$, the diagonal Fisher information $M_{kk}$, the radiometer SNR, and the characteristic strain $h_c(f,\hat{\Omega})$) are all modulated by the array's angular response, not by the injected signal alone. Figure~\ref{fig:snr_hd_single_hotspot} shows the array's response to a hotspot with concentration $\kappa=10$ and isotropic floor $f_{\mathrm{iso}}=0.10$, placed at right ascension and declination $(\mathrm{RA},\mathrm{Dec})=(180^\circ,0^\circ)$. The expected SNR map (panel a) peaks at the injected location but is mildly reshaped by the surrounding pulsar distribution, and the anisotropic injection registers as a deviation of the per-pair ORF from the isotropic Hellings-Downs curve (panel b). How strongly a hotspot shows up depends on where it falls relative to the array's sensitive regions, which we examine directly in the two-hotspot case below. Table~\ref{tab:fiso_snr} reports the total detection SNR for this configuration as the isotropic floor $f_{\mathrm{iso}}$ is varied. The full-Fisher total exceeds the radiometer at every $f_{\mathrm{iso}}$, including the pure-hotspot limit ($f_{\mathrm{iso}}=0$), and as $f_{\mathrm{iso}}\to1$ the full-Fisher total recovers the standard \texttt{hasasia} detection SNR. This validates the injection machinery against the isotropic limit and quantifies the coherence gap between the radiometer and full-Fisher totals (\S\ref{subsec:total_snr}).

\begin{table}[htb]
 \caption{\label{tab:fiso_snr} Total detection SNR for a single von Mises-Fisher hotspot ($\kappa=10$), at $(\mathrm{RA},\mathrm{Dec})=(180^\circ,0^\circ)$, with an injected GWB ($A_{\mathrm{GWB}}=2.4\times10^{-15}$, $\gamma=13/3$) for the 16-year SPTA, as a function of the isotropic-floor fraction $f_{\mathrm{iso}}$. The radiometer and full-Fisher totals are computed from Eqs.~\eqref{eqn:snr_rad} and~\eqref{eqn:snr_full} respectively. The full-Fisher total exceeds the radiometer even for a pure hotspot ($f_{\mathrm{iso}}=0$), and the gap widens monotonically as the hotspot becomes more diffuse toward the isotropic limit ($f_{\mathrm{iso}}=1$), where the full-Fisher total recovers the standard isotropic \texttt{hasasia} detection SNR.}
 \begin{center}
 \begin{tabular}{cccc}
 \br
 $f_{\mathrm{iso}}$ & $\mathrm{SNR}_{\mathrm{rad}}$ & $\mathrm{SNR}_{\mathrm{full}}$ & $\mathrm{SNR}_{\mathrm{full}}/\mathrm{SNR}_{\mathrm{rad}}$ \\
 \mr 
 0.0 & 1.47 & 17.92 & 12.2 \\
 0.1 & 1.34 & 16.83 & 12.6 \\
 0.2 & 1.21 & 15.79 & 13.1 \\
 0.3 & 1.08 & 14.79 & 13.7 \\
 0.4 & 0.96 & 13.85 & 14.5 \\
 0.5 & 0.84 & 12.97 & 15.5 \\
 0.6 & 0.72 & 12.18 & 16.9 \\
 0.7 & 0.62 & 11.49 & 18.6 \\
 0.8 & 0.53 & 10.92 & 20.6 \\
 0.9 & 0.47 & 10.50 & 22.4 \\
 1.0 & 0.44 & 10.22 & 23.1 \\
 \br
\end{tabular}
\end{center}\end{table} 

\subsubsection{Two Hotspots}

\begin{figure*}
    \centering
    \begin{subfigure}[t]{0.49\textwidth}
        \centering
        \includegraphics[width=\textwidth]{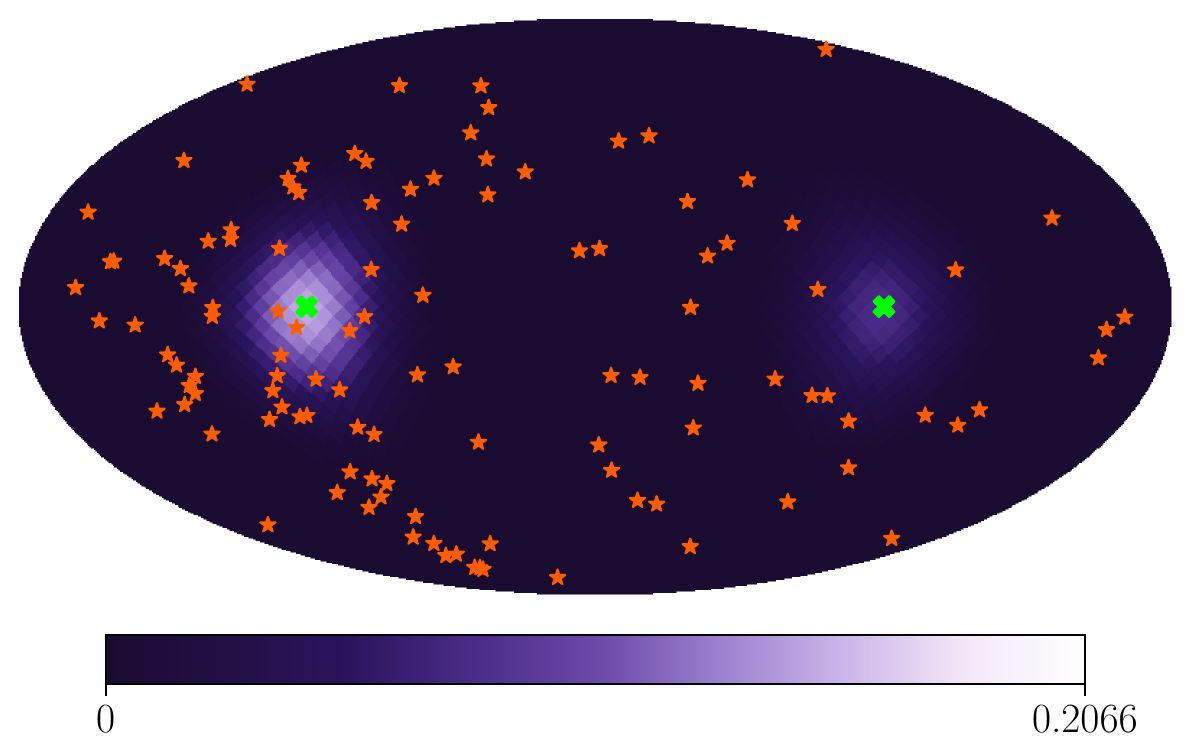}
        \caption{Symmetric}
        \label{fig:two_hotspot_sym}
    \end{subfigure}
    \begin{subfigure}[t]{0.49\textwidth}
        \centering
        \includegraphics[width=\textwidth]{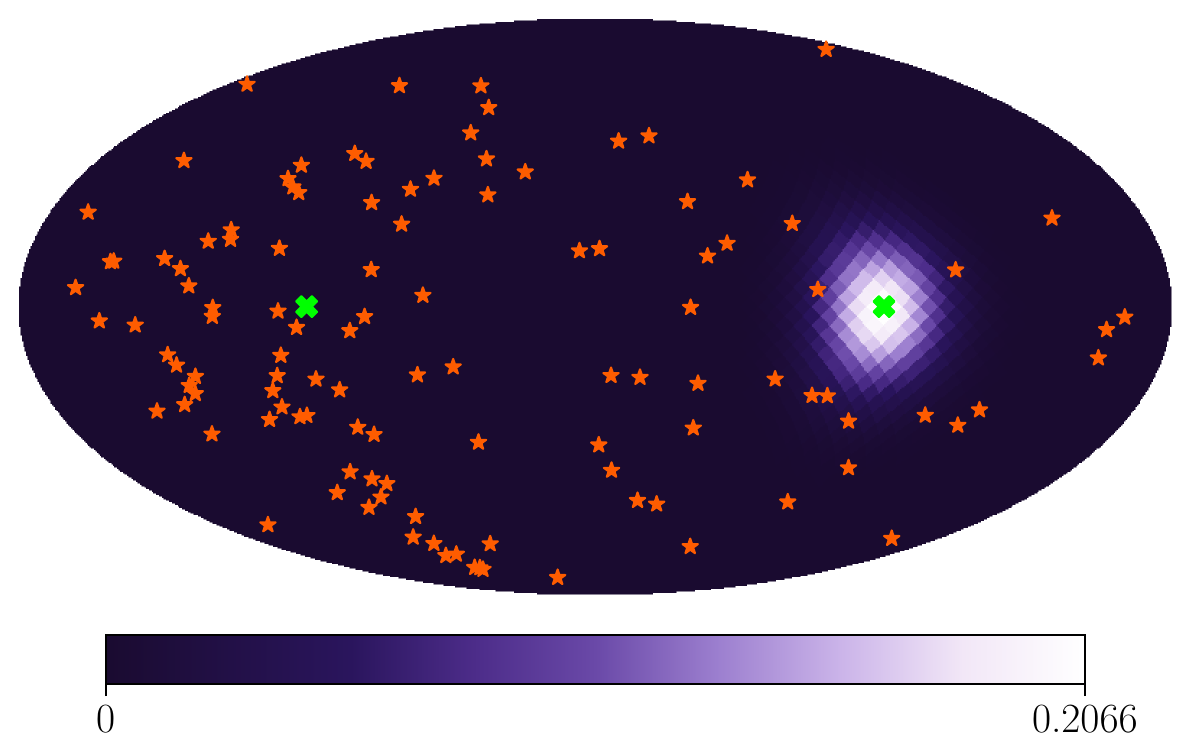} 
        \caption{Asymmetric}
        \label{fig:two_hotspot_asym}
    \end{subfigure}
    \caption{Per-pixel radiometer detection SNR (Eq.~\eqref{eqn:snr_perpix}) for two injected hotspots at $\mathrm{RA}=90^\circ$ (pulsar-sparse, 5 pulsars within $30^\circ$) and $\mathrm{RA}=270^\circ$ (pulsar-dense, 16 within $30^\circ$), both at $\mathrm{Dec}=0^\circ$, with an injected GWB ($A_{\mathrm{GWB}}=2.4\times10^{-15}$, $\gamma=13/3$, $f_{\mathrm{iso}}=0.05$) for the 16-year SPTA. Green crosses mark the injected hotspot centers. (a) Symmetric: equal amplitudes and $\kappa=15$. Despite the identical injection, the hotspot in the pulsar-dense region ($\mathrm{RA}=270^\circ$) reaches roughly twice the SNR of its counterpart in the pulsar-sparse region, isolating the array's angular response. (b) Asymmetric: amplitude $1.0$, $\kappa=20$ at $\mathrm{RA}=90^\circ$ (pulsar-sparse) and amplitude $0.5$, $\kappa=10$ at $\mathrm{RA}=270^\circ$ (pulsar-dense). The stronger, more concentrated hotspot dominates the map despite its pulsar-sparse placement, while the weaker hotspot in the dense region is barely registered. This shows that intrinsic hotspot strength can outweigh the array's angular response. Both panels show the same color scale. Total detection SNR (radiometer / full-Fisher): symmetric $1.22/13.84$, asymmetric $1.31/12.75$.}
    \label{fig:two_hotspot}
\end{figure*}

For the two-hotspot case, we inject two equal hotspots, one at $\mathrm{RA}=90^\circ$ (a pulsar-sparse region) and one at $\mathrm{RA}=270^\circ$ (a pulsar-dense region), both at $\mathrm{Dec}=0^\circ$, on top of an injected GWB ($A_{\mathrm{GWB}}=2.4\times10^{-15}$, $\gamma=13/3$) with an isotropic floor $f_{\mathrm{iso}}=0.05$, shown in Figure~\ref{fig:two_hotspot}. In the symmetric configuration (equal amplitude, $\kappa=15$), the two hotspots are injected identically, so the difference in their expected SNR is due entirely to the array's angular response: the hotspot in the pulsar-dense region reaches roughly twice the SNR of the one in the pulsar-sparse region. The asymmetric configuration (amplitudes $1.0$ and $0.5$, $\kappa=20$ and $\kappa=10$) places the stronger hotspot in the pulsar-sparse region; it nonetheless dominates the map, showing that intrinsic hotspot brightness can outweigh where it falls relative to the array.  

\subsection{Forecasting Directional Sensitivity}

Looking forward, we can use the same methods to forecast the SPTA to the 40-year configuration. The 16-year slice of the SPTA has 115 pulsars and an isotropic background SNR of $\sim 7$ and the 40-year slice of the SPTA has 157 pulsars and an isotropic background SNR of $\sim 32$. 

\subsubsection{Sensitivity Skymaps}

\begin{figure*}
    \centering
    \begin{subfigure}[t]{0.49\textwidth}
        \centering
        \includegraphics[width=\textwidth]{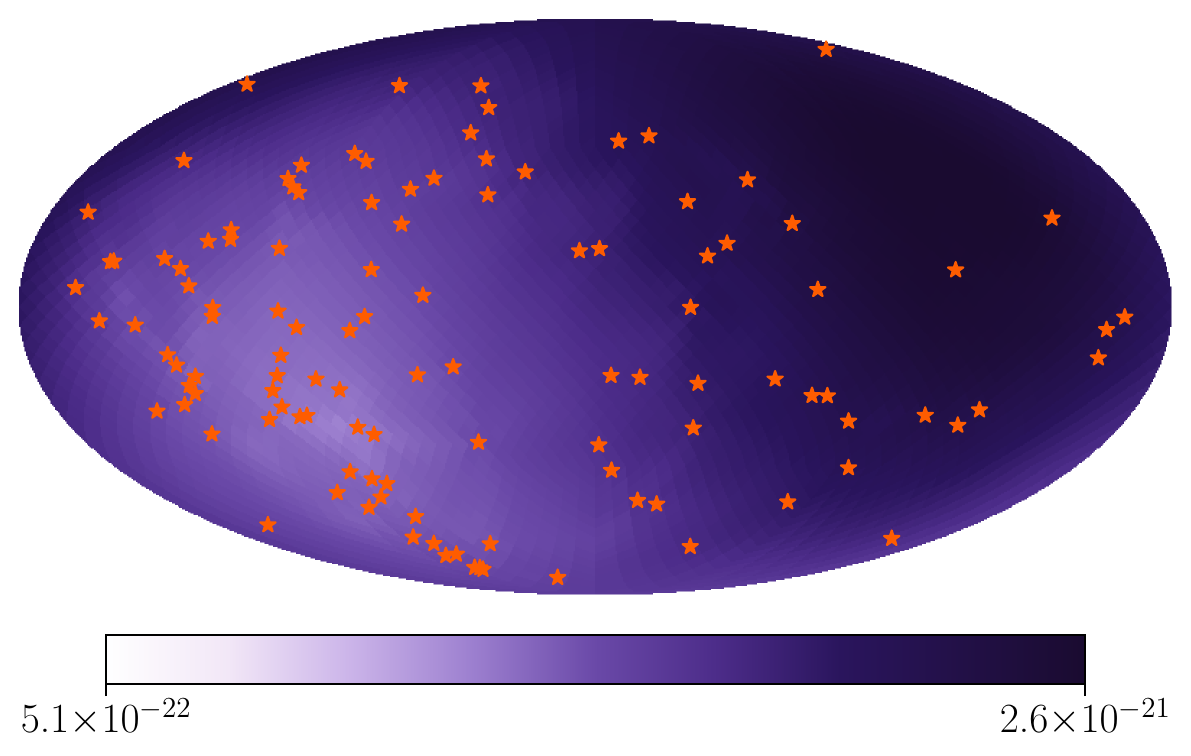}
        \caption{16-year SPTA}
        \label{fig:16yr_seff_skymap}
    \end{subfigure}
    \begin{subfigure}[t]{0.49\textwidth}
        \centering
        \includegraphics[width=\textwidth]{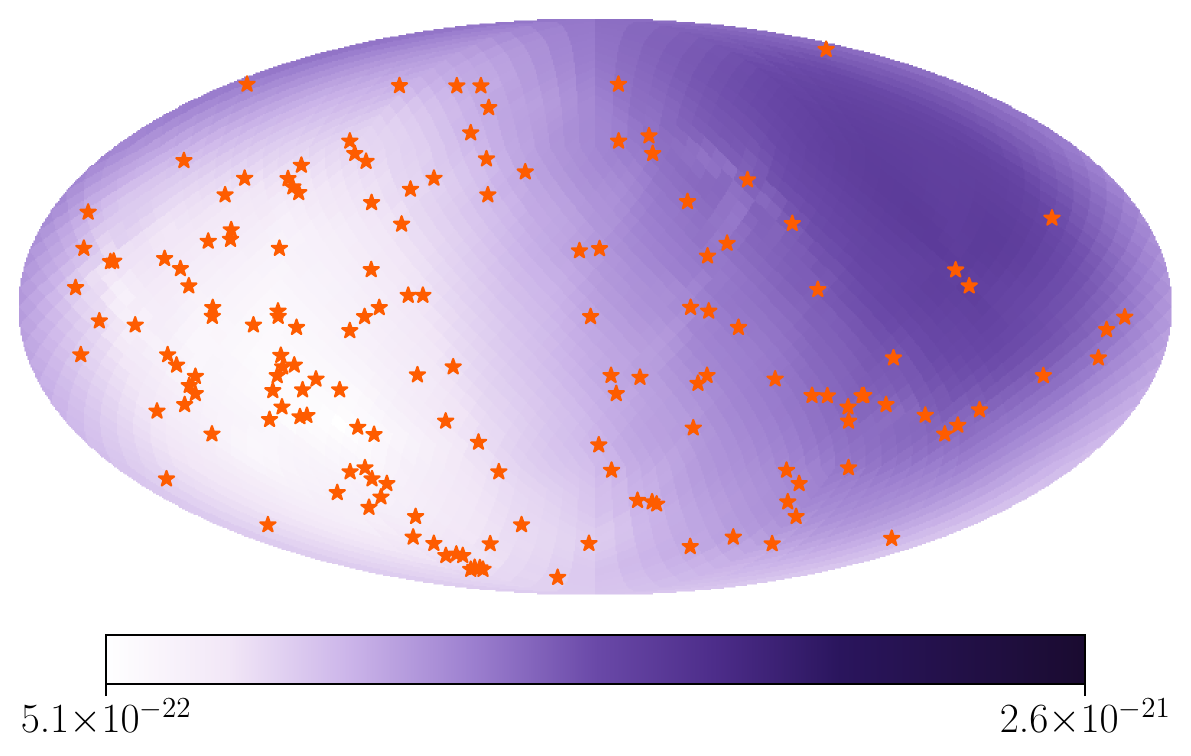} 
        \caption{40-year SPTA}
        \label{fig:40yr_seff_skymap}
    \end{subfigure}
    \caption{Per-pixel radiometer effective sensitivity $S_{\mathrm{eff}}^{\mathrm{rad}}$ at the 16-year SPTA's peak sensitivity frequency. (a) shows the 16-year SPTA with 115 pulsars and (b) shows the 40-year SPTA with 157 pulsars. Note that the color bar bounds are fixed between the two figures to highlight the difference in sensitivity.}
    \label{fig:forecast_seff_skymap}
\end{figure*}

We begin by comparing the effective sensitivity at the peak sensitivity frequency of the 16-year SPTA. Figure~\ref{fig:forecast_seff_skymap} shows the per-pixel radiometer effective sensitivity  $S_{\mathrm{eff}}^{\mathrm{rad}}$ (Eq.~\eqref{eqn:seff_pix_rad}) for the 16-year (a) and 40-year (b) configurations, with the colorbar bounds held fixed between panels so the two are directly comparable. Extending the array to the 40-year configuration lowers $S_{\mathrm{eff}}$ across the entire sky, improving the median per-pixel sensitivity by $\sim 1.9$. The improvement is not spatially uniform with the largest gains appearing in directions where the newly added pulsars fall, so the forecast both lowers the overall sensitivity floor and reshapes the array's angular response as the sky coverage fills in. 

\begin{figure*}
    \centering
    \begin{subfigure}[t]{0.49\textwidth}
    \centering
    \includegraphics[width=\textwidth]{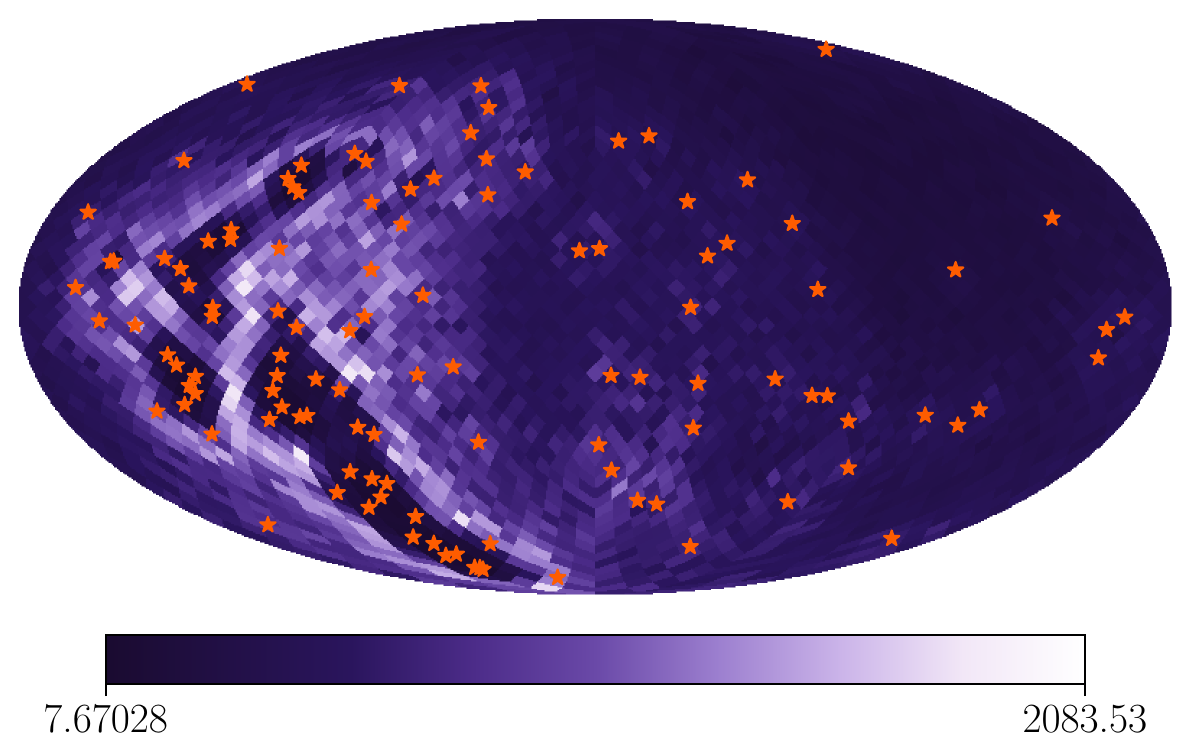}
        \caption{16-year SPTA}
        \label{fig:ratio_16yr}
    \end{subfigure}
    \begin{subfigure}[t]{0.49\textwidth}
    \centering
    \includegraphics[width=\textwidth]{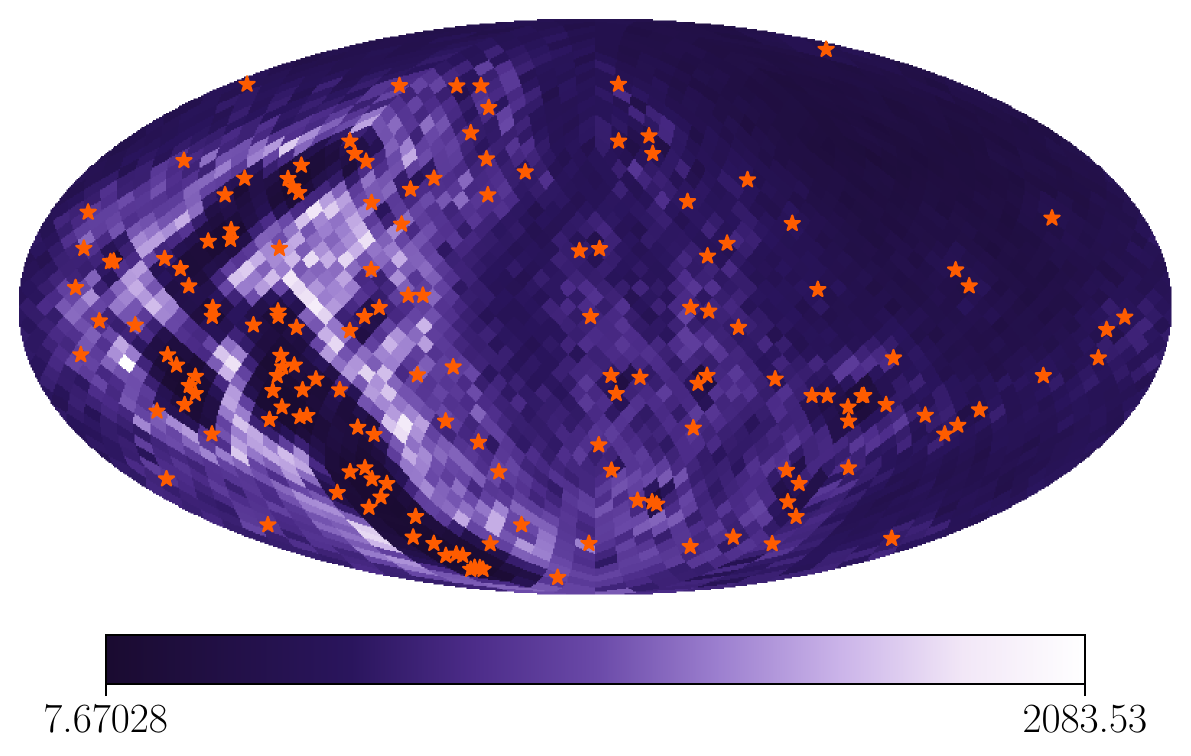} 
        \caption{40-year SPTA}
        \label{fig:ratio_40yr}
    \end{subfigure}
    \caption{Per-pixel ratio of the full-Fisher to the radiometer effective sensitivity, $S_{\mathrm{eff}}^{\mathrm{full}}/S_{\mathrm{eff}}^{\mathrm{rad}} = \sqrt{[M^{-1}]_{kk}}\big/\big(1/\sqrt{M_{kk}}\big)$, at the 16-year SPTA's peak sensitivity frequency for the 16-year (a) and 40-year (b) SPTA. The ratio is independent of $T_{\mathrm{obs}}$ (it cancels), so it isolates the geometric off-diagonal couplings. Both panels share the same colorbar for direct comparison.}
\label{fig:forecast_ratio_fisher_skymap}
\end{figure*}

Figure~\ref{fig:forecast_ratio_fisher_skymap} compares the per-pixel ratio of the full-Fisher to the radiometer effective sensitivity for the two configurations, evaluated at the peak sensitivity frequency. This ratio measures the size of the Cauchy-Schwarz gap of \S\ref{subsec:fisher_seff}, showing how much directional information the off-diagonal Fisher couplings carry, and equivalently quantifies the array's over-parameterization relative to its $\sim\ell_{\mathrm{eff}}^2$ independently constrained sky modes. The two maps look very similar by eye, and that is the point. Because the ratio is independent of $T_{\mathrm{obs}}$ (it cancels), it is a purely geometric quantity set by the pulsar sky distribution rather than integration time. The added baseline therefore drops out of the ratio entirely, and the small difference between the two maps comes from the growth in pulsar population, from 115 to 157 pulsars. Even so, the effect is modest, with the angular pattern and overall magnitude of the ratio nearly unchanged. The angular pattern of the radiometer/full-Fisher gap is therefore a structural feature of the array geometry. Observing longer drives the sensitivity curves themselves down, but it does not close the gap between the two estimators. The absolute scale of the ratio is set by the truncated-SVD regularization of the near-singular full-Fisher inverse, which we characterize in \ref{app:regularization}, while the near-invariance of the pattern across configurations is robust to that choice.

\subsubsection{Directional Sensitivity Curves}

\begin{figure}
    \centering
    \includegraphics[width=0.6\linewidth]{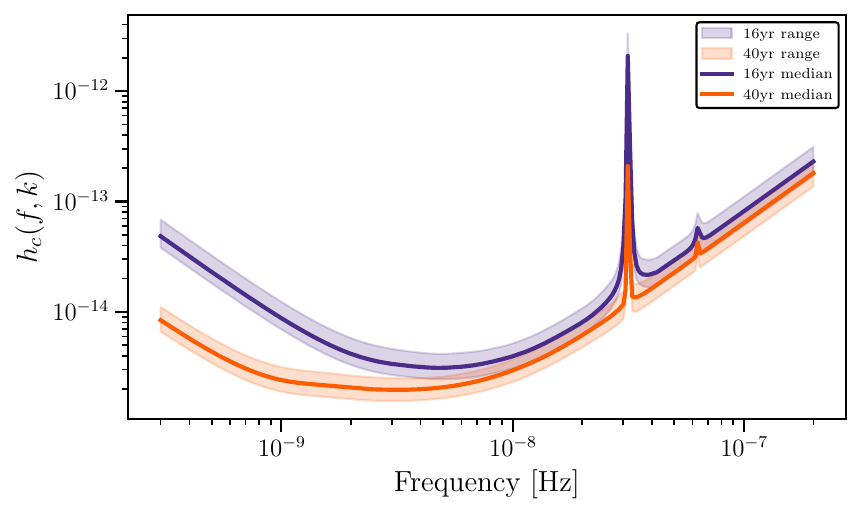}
    \caption{Per-pixel characteristic strain sensitivity curves for the 16-year SPTA (purple) and the 40-year SPTA (orange), with the full range of the curves plotted along with the median sensitivity.}
    \label{fig:forecast_hc_perpixel}
\end{figure}

Beyond the single-frequency skymaps, the forecast carries through to the full directional sensitivity curves. Figure~\ref{fig:forecast_hc_perpixel} shows the characteristic strain curves $h_c(f,\hat{\Omega})$ for the two configurations, with the shaded band spanning best to worst pixel and the solid line showing the median. The largest gain is at low frequencies, where the longer observing baseline extends the sensitive band and the added pulsars deepen it, consistent with the expected scaling for a growing array \cite{siemens_stochastic_2013}. Our forecast adds both pulsars and observing time; the dominant low-frequency gain is consistent with Moursy et al. (2026) \cite{moursy_anisotropy_2026}, who isolate the observing baseline and find its effect concentrated at the lowest frequencies.

\begin{figure}
    \centering
    \includegraphics[width=0.6\linewidth]{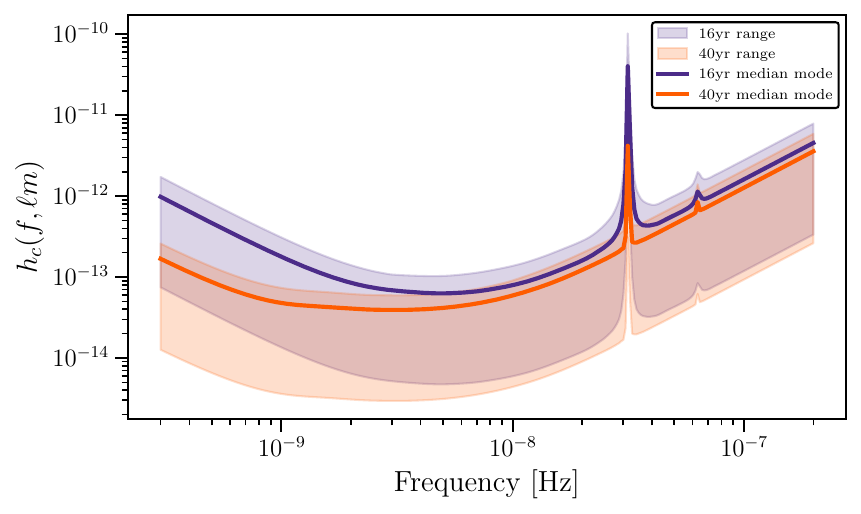}
    \caption{Per-mode characteristic-strain sensitivity curves in the spherical harmonic basis for the 16-year (purple) and 40-year (orange) SPTA, showing the best, median, and worst $(\ell,m)$ modes ranked at the 16-year SPTA's peak sensitivity frequency. Both configurations use the same basis and the same number of modes ($\ell_{\max}=47$), so the comparison isolates the effect of the larger array. The monopole ($\ell=0,m=0$) is the most sensitive in both; the least-sensitive modes are $(\ell=44,m=-44)$ and $(\ell=46,m=1)$ respectively.}
    \label{fig:forecast_permode}
\end{figure}

Figure~\ref{fig:forecast_permode} shows the same forecast in the spherical-harmonic basis, per mode. The monopole ($\ell=0,m=0$) remains the most sensitive mode and recovers the isotropic limit. The forecast improves every mode with the added pulsars tightening the constraint on the higher $\ell$ modes that the 16-year array constrains only weakly.

\begin{figure}
    \centering
    \includegraphics[width=0.6\linewidth]{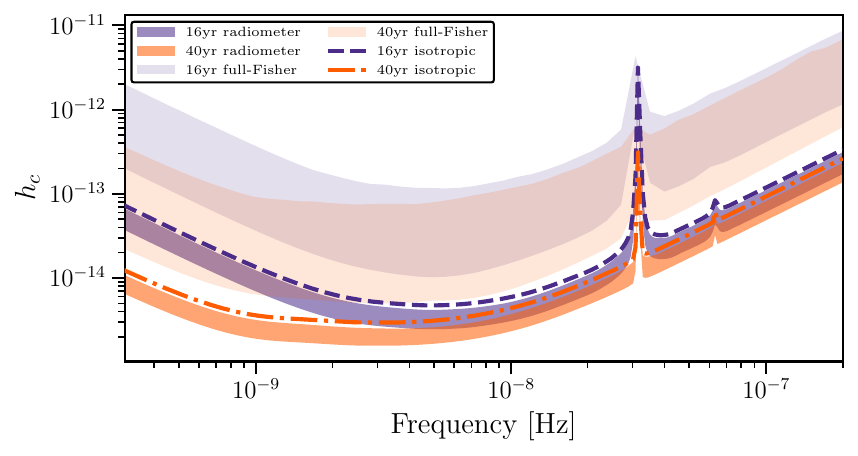}
    \caption{Forecast of the per-pixel characteristic strain sensitivity for the 16-year (purple) and 40-year (orange) SPTA, with the isotropic curve bracketed by the directional estimators in each configuration. For each array, the lower (more opaque) band is the radiometer per-pixel envelope $h_c^{\mathrm{rad}}$, the upper (fainter) band is the full-Fisher envelope $h_c^{\mathrm{full}}$, and the line threading between them is the isotropic \texttt{hasasia} curve (16-year dashed, 40-year dash-dot). The directional Cauchy-Schwarz ordering holds in both configurations. The larger array shifts every band and curve downward.}
    \label{fig:forecast_radio_full_isotropic}
\end{figure}

Finally, Figure~\ref{fig:forecast_radio_full_isotropic} brackets the isotropic sensitivity curve by the radiometer and full-Fisher directional envelopes for both configurations. The bracket persists in each, and the larger array shifts the whole structure downward.

\section{Conclusion}\label{sec:conclusion}

We have developed an analytical framework for directional anisotropic sensitivity curves in pulsar timing arrays, implemented in \texttt{hasasia}. Starting from the cross-correlation estimator, we recast the directional Fisher information as effective sensitivity curves on the sky, $S_{\mathrm{eff}}(f,\hat\Omega)$ and $h_c(f,\hat\Omega)$, and carried two estimators throughout: the radiometer, which treats each direction as an isolated point source, and the full-Fisher, which marginalizes each direction over the rest of the sky. The two are ordered by a Cauchy-Schwarz inequality whose angular pattern measures how strongly general relativity's quadrupolar antenna pattern couples nearby sky directions \cite{vallisneri_use_2008}. We implemented the framework in five sky-decomposition bases on a common Fisher backbone, pixel, spherical harmonic, square-root spherical harmonic, radiometer, and principal-map, the last being the array's own eigenbasis, where the radiometer and full-Fisher estimators coincide. 

We demonstrated the framework on a simulated IPTA-like array, used as a worked example rather than an object of study: characterizing the array's angular response, dissecting the inter-pixel and inter-multipole structure of the Fisher matrix, quantifying the detectability of injected von Mises-Fisher hotspots, and forecasting sensitivity gain from a 16-year to a 40-year configuration. A recurring theme is that the directional quantities we report are shaped by the array's angular response as much as by the injected signal, so a quantitative model of that response, which this framework provides, is a prerequisite for separating the astrophysical anisotropy of the background from the instrumental anisotropy of a non-uniform array. 

Our framework shares its formal core with the anisotropic Fisher analyses of Ali-Ha\"imoud et al. (2020, 2021) \cite{ali-haimoud_fisher_2020, ali-haimoud_insights_2021} and Pol et al. (2022) \cite{pol_forecasting_2022}: the directional Fisher matrix, the weak-signal limit, and the dirty/clean map-making machinery, including the principal-map basis of \cite{ali-haimoud_insights_2021}. We differ in three ways. First, where those studies forecast detection statistics and upper limits, we produce directional effective sensitivity curves $S_\mathrm{eff}(f,\hat\Omega)$ and $h_c(f,\hat\Omega)$, extending the Hazboun, Romano, \& Smith (2019) \cite{hazboun_realistic_2019} and \texttt{hasasia} \cite{hazboun_hasasia_2019} sensitivity curve formalism to the anisotropic case. Second, we report the radiometer and full-Fisher estimators as a Cauchy-Schwarz-ordered pair across all our bases, rather than selecting one. Third, the framework is released as a documented, maintained extension to \texttt{hasasia} wired to per-pulsar noise models, usable for forecasting arbitrary array configurations. In assembling it we have also collected the directional Fisher, map-making, and basis-decomposition results from across the PTA anisotropy literature into a single consistent notation, which we hope is useful in its own right.

Several directions follow naturally. Because the background's anisotropy is expected to grow with frequency \cite{sato-polito_exploring_2024} and a hotspot may be produced by a single loud binary or by several quieter binaries close together on the sky \cite{mingarelli_local_2017}, directional sensitivity curves provide a bridge between stochastic-background and continuous-wave searches, quantifying direction by direction and frequency by frequency where a given array is best placed to detect an emerging hotspot. The implementation already admits a frequency-dependent angular power $P(f, \hat{\Omega})$, and the natural next step is to apply the framework to real PTA datasets. Doing so will also mean choosing the basis truncation with care, since setting $\ell_{\mathrm{max}}$ below the array's angular resolution leaks small-scale power into the retained modes and can bias the recovered anisotropy \cite{semenzato_bias_2025,agarwal_addressing_2026}. We have worked throughout in the weak-signal, Earth-term cross-correlation limit, and the pixel-basis inversions are regularized (see \ref{app:regularization}) to reflect that an array constrains only its $\sim\ell_{\mathrm{eff}}^2$ resolvable modes; relaxing these assumptions is left to future work. 

\section*{Acknowledgements}

We thank Deepali Agarwal, Yacine Ali-Ha\"imoud, Taha Moursy, and Joe Romano for their thorough feedback and useful discussions.

DJO and JSH's work was supported by NSF Physics Frontiers Center award No. 2020265 and No. 2607948. JGB and JSH acknowledge support from NSF CAREER Award No. 2339728. JGB is supported in part through NASA and Oregon Space Grant Consortium, cooperative agreement 80NSSC20M0035. JSH is also supported through an Oregon State University start-up fund. \par
\textit{Software:} \texttt{hasasia \cite{hazboun_hasasia_2019}}, \texttt{SciPy} \cite{scipy}, \texttt{NumPy} \cite{numpy}, \texttt{Matplotlib} \cite{matplotlib}, \texttt{HEALPix} \cite{gorski_healpix_2005}, \texttt{healpy} \cite{zonca+2019+healpy}.\par

\clearpage

\appendix

\section{Regularization of the full-Fisher inversion}\label{app:regularization}

Computing the full-Fisher effective sensitivity (Eq.~\eqref{eqn:seff_pix_full}) and the clean-map uncertainty (Eq.~\eqref{eqn:clean_sigma}) requires inverting the per-frequency pixel Fisher matrix $M(f)$, which is severely ill-conditioned. As discussed in \S\ref{sec:results}, the choice $N_{\mathrm{side}}=16$ keeps the pixel count below the pair count ($N_{\mathrm{pix}}=3072 < N_{\mathrm{pair}}=6555$ for the $16$-year array), so the near-singularity is not a rank deficiency. It comes instead from the quadrupolar antenna response, which constrains only $\sim\ell_{\mathrm{eff}}^2$ independent sky modes and leaves the remaining eigenvalues to decay steeply. This is not an artifact of working frequency by frequency, since the conditioning is set by the geometry of the pairwise response, which is frequency independent, so we expect a time-domain treatment to face the same near-null modes. The inversion is therefore a regularized numerical operation rather than an analytic one, and it only affects the full-Fisher branch, since the radiometer uses the diagonal of $M$ alone (Eq.~\eqref{eqn:seff_pix_rad}) and never inverts it.

\begin{figure}[h!]
    \centering
    \includegraphics[width=0.6\linewidth]{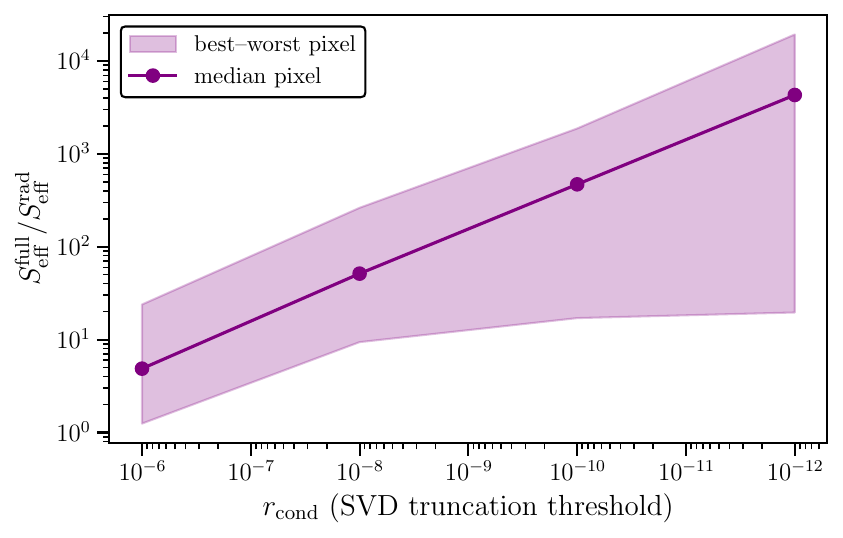}
    \caption{Dependence of the per-pixel full-Fisher to radiometer effective-sensitivity ratio $S_{\mathrm{eff}}^{\mathrm{full}}/S_{\mathrm{eff}}^{\mathrm{rad}}$ on the truncated-SVD threshold $r_{\mathrm{cond}}$, for the 16-year SPTA at the peak sensitivity frequency. The solid line is the median pixel and the shaded band spans the best to worst pixel. The ratio grows by more than two orders of magnitude as the threshold is lowered, confirming that its absolute scale is set by the regularization. Our fiducial choice is $r_{\mathrm{cond}}=10^{-10}$.}
    \label{fig:rcond_sweep}
\end{figure}

We regularize by a truncated singular value decomposition of $M(f)$. Because $M(f)$ is symmetric and positive semi-definite, its singular values are its eigenvalues $\lambda_n$, and we keep $1/\lambda_n$ only where $\lambda_n > r_{\mathrm{cond}}\lambda_{\mathrm{max}}$, discarding the rest. Our fiducial threshold $r_{\mathrm{cond}}=10^{-10}$ sits several orders of magnitude above the double-precision floor, so the retained modes are numerically meaningful rather than dominated by roundoff. Lowering it toward that floor does not converge to a regularization-free answer: once $\lambda_n/\lambda_{\mathrm{max}}$ reaches $\sim10^{-16}$, the eigenvalues are set by roundoff rather than by the array, so inverting them amplifies numerical noise instead of adding sky information, and the ratio in Figure~\ref{fig:rcond_sweep} would continue to grow without ever stabilizing. The threshold is otherwise not privileged, since the eigenvalue spectrum decays smoothly rather than showing a clear gap between constrained and unconstrained modes, so any single choice is a convention. These $\lambda_n$ are the Fisher eigenvalues of the basis's own principal-map decomposition (\S\ref{sec:bases}), so each basis regularizes its own matrix rather than a shared spectrum, and the near-null modes we discard are the least-detectable principal maps.

Figure~\ref{fig:rcond_sweep} shows why this matters. As $r_{\mathrm{cond}}$ is lowered from $10^{-6}$ to $10^{-12}$, the median ratio $S_{\mathrm{eff}}^{\mathrm{full}}/S_{\mathrm{eff}}^{\mathrm{rad}}$ grows by more than two orders of magnitude, so the absolute scale of the full-Fisher quantities is set by the truncation and not by the array alone. Lowering the threshold keeps more modes, which sounds like adding information, but each retained near-null mode contributes a term proportional to $1/\lambda_n$ to the marginal variance $[M^{-1}]_{kk}$. Keeping more poorly measured modes therefore raises $S_{\mathrm{eff}}^{\mathrm{full}}$ at every pixel, since each direction's uncertainty now includes the freedom of more barely constrained sky patterns. The total detection SNR of Eq.~\eqref{eqn:snr_full} is not built from $S_{\mathrm{eff}}^{\mathrm{full}}$ at all. It applies $M$ forward and never inverts it, so it is independent of the threshold. The rising $S_{\mathrm{eff}}^{\mathrm{full}}$ is an estimation error bar, the marginal uncertainty on the power in one direction, not the detection statistic. The qualitative structure, however, is not affected: the Cauchy-Schwarz ordering $S_{\mathrm{eff}}^{\mathrm{full}} \geq S_{\mathrm{eff}}^{\mathrm{rad}}$ holds at every pixel, and the angular pattern of the ratio is preserved throughout. This separates the reported quantities into two classes. The radiometer maps, the isotropic-limit total SNRs (which evaluate $P^TMP$ directly, with no inversion), and any ratio in which $T_{\mathrm{obs}}$ cancels are independent of the truncation. The absolute full-Fisher $S_{\mathrm{eff}}^{\mathrm{full}}$ and $h_{\mathrm{c}}^{\mathrm{full}}$, and the size of their ratio to the radiometer, do depend on it. What this means is that the absolute full-Fisher sensitivities are set by the regularization rather than by the array alone, and we report them at $r_{\mathrm{cond}}=10^{-10}$ to illustrate the estimator's behavior rather than as figures of merit for the array. This does not weaken the conclusions of the paper, because none of them rest on those absolute values. The Cauchy-Schwarz ordering holds at every threshold, the angular pattern of the gap is preserved across the full $r_{\mathrm{cond}}$ sweep, and the isotropic-limit checks and total detection SNRs never invert $M$ at all. The substantive results of this paper therefore rest entirely on the regularization-independent quantities.

\clearpage

\bibliography{aniso}


\end{document}